 \documentclass[12pt]{article}
 \usepackage{verbatim} 
\usepackage{latexsym}
\usepackage{amsmath}
\usepackage{graphics}
\usepackage{epsfig}
\usepackage{graphicx}
\usepackage{fancybox} 
 \usepackage{ifsym} 
 \usepackage{multirow,bigdelim} 

 \usepackage[cmtip,arrow]{xy} 
 \usepackage{pb-diagram,pb-xy} 
 \usepackage{pb-diagram,lamsarrow,pb-lams} 

 \usepackage{cronoga} 
 \usepackage{color}

  \newif\iflatextortf  

 \iflatextortf  
   {} 
 \else 
   \makeatletter 
   \renewenvironment{quotation}  
                 {\list{}{\listparindent=10pt
                          \itemindent    \listparindent 
                          \leftmargin=15pt
                          \rightmargin=15pt
                          \topsep=2pt
                          \parsep        \z@ \@plus\p@}%
                  \item\relax} 
                 {\endlist} 
   \makeatother 
 \fi 

 \renewcommand{\baselinestretch}{1.0}
 \def\rr{\noindent - }

\def\ma1{\!+\!1} 
\def\me1{\!-\!1}

\def\myqt{\sl  }   

\def\argmin \mathop{\rm argmin}
\def\Re{I\kern -0.37 em R}
\def\Na{I\kern -0.37 em N}
\def\Qe{I\kern -0.37 em Q}

\def\Chi2{\mbox{Q}}

\def\Re{{R}}

 \definecolor{myblue}{rgb}{0.1,0.1,0.6} 
 \definecolor{myred}{rgb}{0.8,0.1,0.1} 
\definecolor{mygreen}{rgb}{0.8,0.1,0.1} 
 \def\meuve{}  
 \def\meugr{}  

\begin{document}



 \title{Jacob's Ladder and              
        Scientific Ontologies}          

 \author{Julio Michael Stern\thanks{ 
   e-mail: {\it jstern\_at\_ime.usp.br} \   
   ulr:  {\it http://www.ime.usp.br/$\sim$jstern}     
   \mbox{} \hspace{55mm} \mbox{}   
   IME-USP -- Institute of Mathematics and Statistics 
   of the University of S\~{a}o Paulo, Brazil.  \hspace{20mm}  \mbox{} 
   Slide presentation at   
   {\it  http://www.ime.usp.br/$\sim$jstern/miscellanea/jmsslide/moebs.pdf}  
   } 
   } 

 \date{Final and corrected version\footnote{ 
    Submitted on January 17, 2013; rev.1 June 02, 2013;  rev.2 June 24, 2013;    
         rev.3 September 12, 2014. 
		Former Appendix B published as Stern,J.M.; Nakano,F. (2014). 
				Optimization Models for Reaction Networks: 
        Information Divergence, Quadratic Programming and Kirchhoff's Laws. 
		    {\it Axioms}, 3, 109-118.}  		
      published at:  \\    
			{\it Cybernetics \& Human Knowing},  2014,  v.21, 3, p.9-43.  
        }

 \maketitle

\mbox{}

  \begin{abstract} 
  The main goal of this article is to use the epistemological 
 framework of a specific version of Cognitive Constructivism   
 to address Piaget's central problem of knowledge construction, 
 namely, the re-equilibration of cognitive structures.  
  The distinctive objective character of this constructivist framework 
	{\meuve is supported by formal inference methods of  
 Bayesian statistics,  and is based on Heinz von Foerster's 
 fundamental metaphor  of   
  {\it objects as tokens for eigen-solutions.} } 
  This epistemological perspective is illustrated using some 
 episodes in the history of chemistry concerning the definition 
 or identification of chemical elements.
  Some of von Foerster's epistemological imperatives 
 provide general guidelines of development and argumentation. 

  \noindent 
 {\meugr {\bf Keywords:} Bayesian statistics; Chemical element; 
  Cognitive constructivism; Development of cognitive structures;   
  Eigen-solution; Objectivity; Ontology alignment; 
  Symbol grounding.}    
                 
 \end{abstract} 

\vfill  

\mbox{}  

\pagebreak 

 \iflatextortf  
  { 
 \else    
  {  \flushright 
 \fi 

  \mbox{} 
   
  {\it 
 An adequate model of knowledge construction must comply with two \\ 
 conditions that are difficult to conciliate: \ To be open to indefinite 
 new\\ possibilities while conserving already constructed cycles of mutual 
 impli-\\ cations  destined to be converted into sub-systems of an expanded  
 system: \\ The issue is therefore to conciliate openness and closure. \\   
  }
    
  Jean Piaget (1976, p.91), 
  commenting von Foerster contribution. 

  \mbox{}

  {\it When a sage points to the moon, 
       only a fool looks at the finger.} \\   
     
   Zen Buddhist proverb.  

\mbox{} 
       
 }

\setcounter{footnote}{1}    

 \section{Introduction} 

   Cognitive Constructivism is based in the general theory of autopoietic 
 (autonomous, self-producing) systems developed by the philosophers 
 Humberto Maturana, Francisco Varela, Heinz von Foerster, Niklas Luhmann, 
 and several others, see Foerster (2003), Maturana and Varela (1980), 
 and Varela (1978). 
  Nevertheless, the specific version of cognitive constructivism 
 {\meuve I use} as the epistemological framework at the core of this article 
 has a very distinctive objective character that, {\meuve I believe}, 
 makes it more suitable for scientific applications.  
  The objective character of this epistemological framework 
	{\meugr is supported by formal inference methods of  Bayesian statistics,   
	and is based on Heinz von Foerster's fundamental metaphor of      
  {\it objects as tokens for eigen-solutions}}\footnote{  \meuve 
 The concept of eigen-solution is the key for an autopoietic system 
to distinguish specific  objects in a cognitive domain. 
 This critical point is further clarified in the following quotations of 
 Heinz von Foerster: 
   
``Objects are tokens for eigen-behaviors. Tokens stand for something else. 
 In exchange for money (a token itself for gold held by one's government, 
 but unfortunately no longer redeemable), tokens are used to gain 
 admittance to the subway or to play pinball machines. 
  In the cognitive realm, objects are the token names we give to our eigen-behavior. 
	 This is the constructivist's insight into what takes place when we talk about 
	 our experience with objects." Segal (2001, p.127).  
   
 ``The meaning of recursion is to run through one's own path again. 
 One of its results is that under certain conditions there exist indeed solutions 
 which, when reentered into the formalism, produce again the same solution. 
 These are called eigen-values, eigen-functions, eigen-behaviors, etc., 
  depending on which domain this formation is applied --- in the domain of numbers, 
	 in functions, in behaviors, etc. The expression `eigen-something' comes from the 
	 German word `self'.  It was coined by David Hilbert in the late 19th century for 
	solutions of problems with a structure very similar to the ones we are talking about.'' 
	Segal (2001, p.145). 
   
  In fact, the expression  coined by Hilbert was not { eigen-something}         
	 but rather, { eigen-solution}  ({\it Eigenl\"{o}sung}). 
	 Since its inception, Hilbert's expression has been widely used in mathematics 
	 and physics. Since the early days of quantum mechanics,  the concept of 
	 eigen-solution has acquired paramount importance in philosophy of science,  
	 where it has been associated to a variety of epistemological metaphors.  
	 This explains my formulation of the metaphor 
	{\it objects are tokens for eigen-solutions}  
	as a generalization directly derived from von Foerster's original formulation 
	{\it objects are tokens for eigen-behaviors}.  }. 
 
  The main goal of this article is to address, according to the 
 cognitive constructivism perspective, some issues related to 
 Piaget's central problem of knowledge construction, as stated at the 
 opening quotations.   
 {\meuve I will} also illustrate this perspective using some 
 episodes in the history of chemistry;  
  Figure 8, at the end of this paper, gives  approximate 
 timelines for the work of some involved scientists. 
  All these episodes concern the discovery, characterization or 
 identification of chemical elements, as they were defined after   
 Lavoisier's chemical revolution. 
 {\meuve  I am} afraid that, in doing these exercises, 
 {\meuve I play} the role of the  fool alluded in the Zen Buddhist 
 aphorism at the opening quotations.  
  Yet, analyzing how objects of scientific knowledge come to be and how 
 it is possible to objectively point at them, that is, understanding 
 scientific understanding,  is an unavoidable role that must be played 
 by those interested in history and philosophy of science. 
 
   The way {\meuve I present} these episodes is certainly not the 
	only way to tell the story. 
   However, the versions {\meuve I present} are, {\meuve I hope}, 
	well grounded in the literature of history of science and its original 
	sources.
   In order to establish this point, {\meuve I made} a conscious effort to
  provide a good assortment of short, clear and pertinent quotations
  from historical documents and influential commentators. 
   Some of these quotations are even self-referential, giving a clear 
 view of how key authors see or evaluate their own work. 
  For example, {\meuve I use} an ``anonymous''  review by Lavoisier 
 of his own work, and also some opinions of Kirchhoff and Balmer 
 concerning the foundations and evolution of spectral analysis, 
 including their own work on the field. 
  These historical examples have been chosen for didactic reasons: 
   They are, {\meuve I hope}, clear and simple, using concepts that are  
  already familiar or easily grasped by a wide audience.  
	{\meuve All examples pertinent to our discussion must involve 
	mathematization, which is necessary for the formulation of invariant 
	quantities of interest.}    
   However, the examples at hand only use pedestrian mathematics.   
     
   It is important to realize that virtually all key foci of natural 
  sciences involve eigen-behaviors, fixed points, invariant quantities, 
  or similar eigen-forms, represented as the most fundamental objects 
  of the pertinent  ontology.  
   For example, in Stern (2011b) {\meuve I present} a formal comparative 
  analysis of the role played by Noether-type theorems in Physics 
  and deFinetti-type theorems in Probability and Statistics.    
   Furthermore, the concept of eigen-solution is amply used  
  in social sciences, ethology, psychology, and many other areas.  
   For example, in economics, prices are often regarded as 
  eigen-values or fixed-points expressing supply and demand 
  equilibria, see Border (1989), Shashkin (1991) 
  and Zangwill and Garcia (1981).    
   For an historical perspective of the {\it invisible hand} 
  responsible for building and maintaining these eigen-solutions,  
  see Ingrao and Israel (1990) and Cerezetti (2013).  
    
 \begin{figure}[b]  
  \iflatextortf
   {} 
  \else  
  \centerline{
  \includegraphics*[height=1.2in]{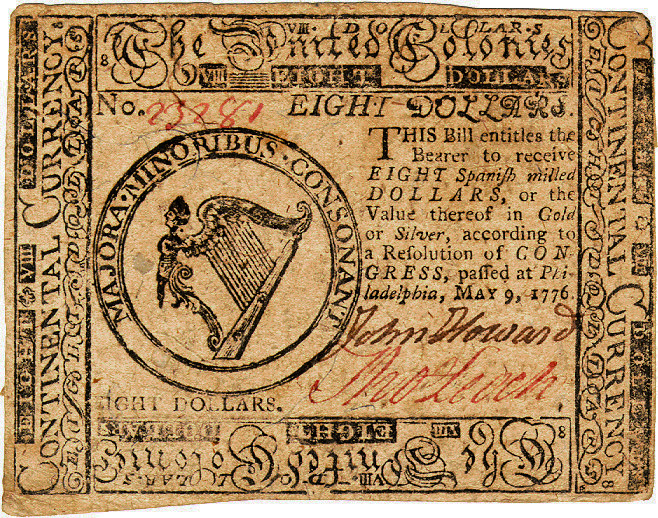} \mbox{} 
  \includegraphics*[height=1.2in]{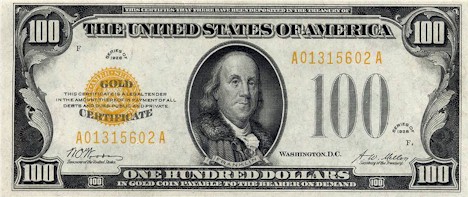}  
  }

  \centerline{
  \includegraphics*[height=1.2in, , width=2.0in]{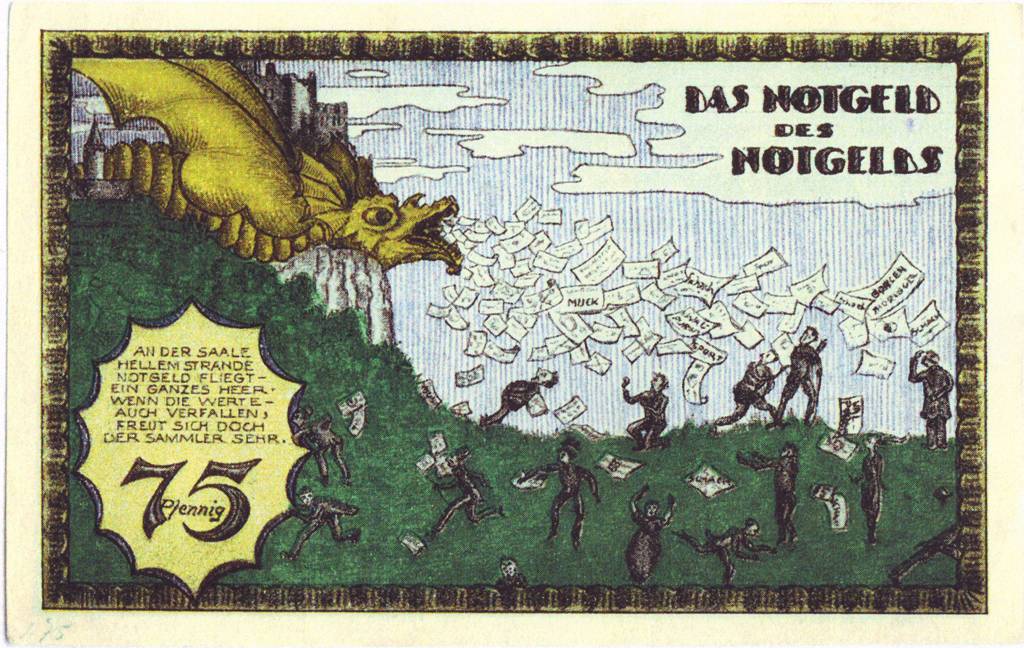} \mbox{}  
  \includegraphics*[height=1.2in, , width=2.0in]{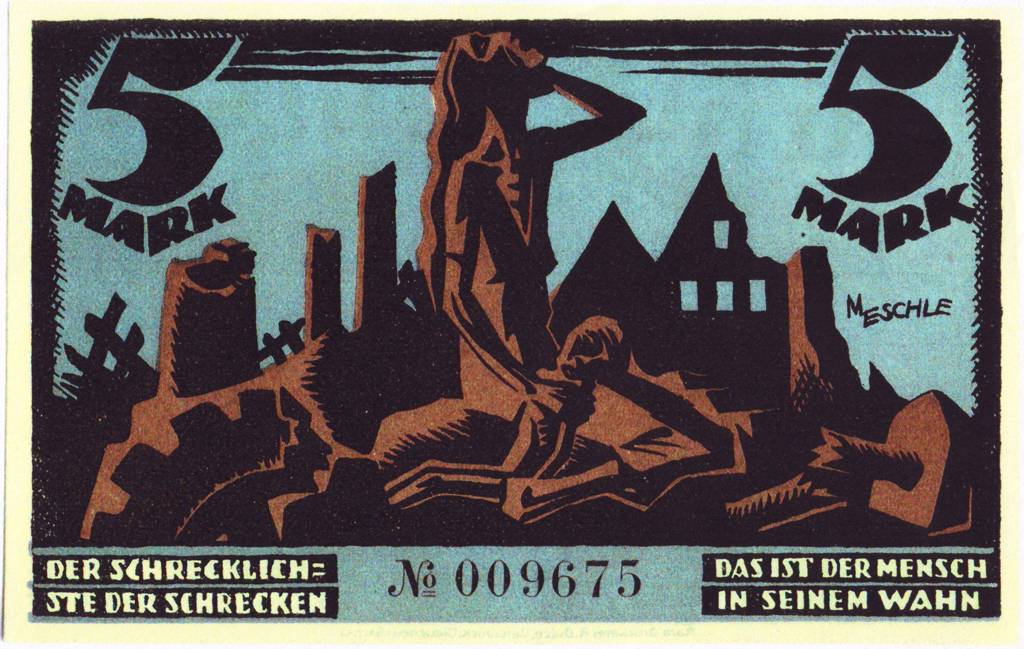} \mbox{}    
  \includegraphics*[height=1.2in, , width=2.0in]{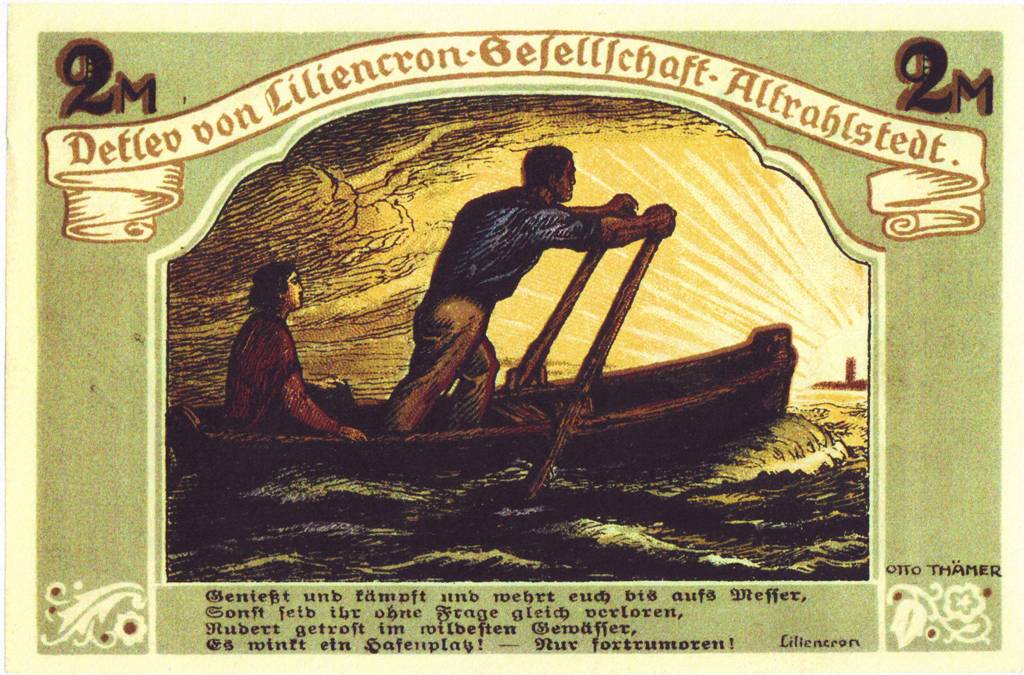}    
  }
 \fi 
 \vspace{0.5mm} 

 \centerline{Figures 1ab: Strongly objective economic tokens;  
                      1cde: Weakly objective economic tokens.} 
 \end{figure}

  Figures 1ab display economic tokens that are strengthen by a 
 chaining process in which they inherit stability and invariance 
 properties of secondary  backup tokens. 
  The US 8-dollar note was issued under the Mint Act of 1792, 
 establishing a dual (alternative) conversion policy at fixed ratios of 
 24.75 grains of gold per dollar, and 15 grains of silver per grain of gold.   
  The US 100-dollar note was issued under the Gold Standard Act of 1900, 
 and was still a convertible (to gold only) currency. 
  After 1971, the US dollar became a free-floating {\it fiat money}.   
  Figures 1cde display some emergency-money cards {\meuve \it (Notgeld)}, 
 weakly objective tokens used at the hyper-inflation period in Germany.
  The 75-pfennig card displays a picture about the inflation process 
 that destabilizes and devaluates economic tokens, plunging 
 the economic system into chaos.     
  The 5-mark card brings the legend   
 ``The most frightful of frights is man in his delusion'', 
 from {\it The Song of the Bell}  by Friedrich Schiller, 
 a reminder of the terrible consequences that may result from the 
 collapse of equilibria and stability conditions that sustain the 
 social fabric of any human civilization.

   Jean Piaget makes his statement at the opening quotations 
 commenting von Foerster contribution in Inhelder (1976, p.91).   
  {\meuve I will} reverse this position using some aphorisms formulated 
 by Heinz von Foerster in search for answers to Piaget's problem.  

The following ordered list of five imperatives suggests a logical concatenation 
of sequential steps in the ontogenesis of an autopoietic system: From the
incapacity to effectively handle some relevant but not yet understandable or even
clearly perceived aspect of the system’s environment; To the struggle to transcend this
blindness or limitation; To the construction of an expanded reality incorporating new objects 
and semantic links; Culminating with re-equilibrations of cognitive structures necessary to
maintain functional coherence and systemic integrity. Of course, several cycles of this
five-step dance can or should be repeated during the system's life or development.

 \begin{enumerate}\iflatextortf{}\else{\itemsep2pt \parskip0pt \parsep0pt}\fi    
 \item 		
     Metaphysical (implicit, negative) imperative: \mbox{} \\       
  {\it Something that cannot be explained cannot be seen.}  \\ 
	(derived from von Foerster, 2003, p. 284) 
	\item 
     Therapeutic Imperative:
  {\it If you want to be yourself, change!}  \\ 
	 (as stated in von Foerster, 2003, p. 303)	
  \item 
		Aesthetical imperative: 
  {\it If you desire to see, learn how to act!} \\ 
	 (as stated in von Foerster, 2003, p. 227)
   \item 
     Ethical imperative: 
  {\it Act so as to increase the number of choices!} \\ 
		(as stated in von Foerster, 2003, p. 227)
   \item 
     Organic imperative: 
  {\it Maintain structural and functional integrity!} \\ 
	(derived from von Foerster, 2003, p. 175)	
  \end{enumerate}  
	

{\meuve Von Foerster's aphorisms condense essential characteristics of the life 
of autonomous organisms or the evolution of abstract autopoietic systems. 
Some of those he designated as {\it imperatives}, also giving them distinctive titles. 
 I took the liberty of using the term {\it imperative} to designate 
all five of von Foerster's aphorisms above, 
also choosing a distinctive title for those that originally did not 
receive this honor. 
 In this way, I have chosen the expressions 
 {\it metaphysical (negative) imperative}   
 and {\it organic imperative} as labels for von Foerster's aphorisms of, respectively, 
 ``something that cannot be explained cannot be seen''  and 
 ``maintain structural and functional integrity''. }    
  
  The (positive) imperatives appearing in this article should not be 
 interpreted as impositive commands. Rather, depending on context or 
 disposition, they can be interpreted as general guidelines for conduct, 
 directions for systemic development, or just as pointers to the natural flow of life. 
  In the same way as the consecrated expression {\it forbidden state}    
  is used in contemporary physics,  a negative imperative should not be 
  interpreted as an arbitrary prohibition, but rather as an indication of 
	existential constraints, excluded states, or impossible 
	pathways\footnote{ 
	 In this way, the Metaphysical (implicit, negative) imperative, namely, 
	Something that cannot be explained cannot be seen!, 
	does not indicate a desire or conscious preference not to see something, 
	but rather the incapacity so to do. 
	Von Foerster (2003, p. 212) illustrates this state of mind with his famous 
	Blind Spot optical effect. Similar blinding effects, known as ambigrams, 
	have been studied by Joseph Jastrow (1899) and Ludwig Wittgenstein (1953). 
	In all these examples, the blinding effect is neither grounded on conscious 
	preferences, nor is it rooted on fundamental physiological limitations, 
	being instead an obstacle that is a consequence of the observer’s 
	current point of view or prior mindset, see Kihlstrom (2006) and Stern (2008b) 
	for further commentaries.
 Overcoming such a blind spot always requires the opportunity to have a 
 transcending point of view, acquiring an innovative insight, or even 
 deeper modifications in the observer’s cognitive structure.}.

  Jacob's Ladder, a classical visual metaphor for evolutionary 
 processes, is depicted in Figures 2a and 2b as a spiral ladder 
 or helical stairway. 
  These two beautiful images will be used in figurative analogies that, 
 {\meuve I hope}, may help to convey the meaning of some  ideas 
 or arguments presented in the text in more abstract format or 
 dryer style.

  Section 2 presents the key concept of eigen-solution and 
 its {\meuve aesthetical} characterization, that is, 
 it explains how we perceive objects in the practice of science. 
  This section illustrates these concepts studying the importance  
 of the scientific equipment developed by Lavoisier and their use 
 in analytical procedures of the new chemistry. 
  {\meuve I shall} also try to understand why these new procedures 
 are perceived  as so powerful, why the objects of the ``new'' chemistry 
 are perceived  as so real, and why some commentators perceive 
 Lavoisier's revolution as a point of transition, from pre-scientific 
 alchemy to scientific chemistry.

   Many traditions consider ethics as {\meuve conformity} to an 
 externally produced (allopoietic) normative regulation, legal code, 
 {\meuve and so forth}. 
  However, there is also a long standing autopoietic tradition  
 in which ethics relates to the effort, endeavor or struggle of an 
 individual, living organism or abstract 
 autopoietic system\footnote{  \meuve 
   The definition of autopoietic system used in this article can be found in 
	Maturana and Valela (1980, Page 78):  
	 ``An autopoietic machine is a machine organized (defined as a unity) 
	  as a network of processes of production (transformation and destruction) 
		 of components which: 
    (i) through their interactions and transformations continuously regenerate 
		and realize the network of processes (relations) that produced them; and 
   (ii) constitute it (the machine) as a concrete unity in space in which they 
	  (the components) exist by specifying the topological domain of its realization 
		as such a network.'' 
       
  This definition aims to capture the essential characteristics of a living system and, 
	 {\meuve I believe},  was extremely successful in achieving its goal. 
	At the same time, this definition is highly abstract: It focus on the organizational 
	structure of systems perceived as living organisms  circumventing, however, 
	the production networks' details of implementation; for example, it does not 
	mention organic chemistry. 
	Hence, the definition of autopoiesis opens the possibility of considering artificial life, 
	social systems, scientific disciplines, and other abstract systems, as sharing some 
	(but not all) of the characteristics of the living organisms of conventional biology, 
	see for example Luhmann (1989) and Krohn et al. (1990). 
	
	Autonomous living-like systems can be ranked by abstraction, as in the following list: 
	Actually known DNA / RNA living organisms; Possible or potentially existing DNA / RNA 
	organisms; Systems based on wet organic chemistry; Systems based on carbon or 
	silicon substrates; Computer simulation of the previous systems; Physically implemented 
	artificial life (AL) systems; Computational AL systems; Social systems; Legal systems.
 Of course, pertinence and order in this list is somewhat arbitrary.

 Maturana and Varela (1980, p.107-109) consider a beehive to be a natural example of 
higher order autopoietic system, suggesting perhaps that social systems should be 
moved cloaser to the head of list above. 
 Even considering someone's favorite ranking order, how far down the list is it appropriate 
to apply the concept of autopoiesis has been the subject of vivid debate. 
 Francisco Varela, Humberto Maturana, and Ricardo Uribe (1974) where among the first to 
study cellular automata as autopoietic systems. However, only a few years later, 
Varela (1978) expressed serious concerns about using the term Autopoiesis for abstract 
systems, suggesting the word Autonomy to be used instead in non-biological contexts. 
 Nevertheless, McMullin and Varela (1997) seem to have found a way of reconciliation 
with at least some of the more general applications of autopoiesis, rediscovering 
computational autopoiesis; see also McMullin (2004).

 Personally, I think that trying to get the genie back in the bottle is an unnecessary and 
even counterproductive effort. 
 In fact, I believe that the case at hand only shows how the best ideas quickly outgrow 
the limits of their originally perceived scope, and how the most powerful concepts soon 
find their way far beyond their first intended applications. 
 Moreover, it has been my personal experience that the integrated study of abstract 
or artificial systems can greatly benefit our understanding of biological systems, 
specially so when investigating difficult questions concerning the always intertwined 
topics of autopoiesis, evolution and self-organization; 
see for example Inhasz and Stern (2010) and the references herein for a concrete 
example concerning the role of genetic introns. 
 Nevertheless, this semantic distinction is not among the core topics of the present 
paper, and the reader should feel free to replace the words 
autopoiesis / autopoietic by autonomy / autonomous whenever he or
she finds it to be appropriate.},   
	 to develop and coherently maintain a basic set of behavioral patterns 
 or core operations, to assemble and stay consistent with kernel procedures, 
 to learn and remember some central concepts or cognitive traits that 
 characterize its (actual, potential or ideal) {\it \meuve modus vivendi}, 
 that is, its form of being or way of existence. 
  Arguably, the most influential work in such an ethics of autonomy
 in mainstream philosophy is that of 
  Baruch Spinoza (1677, see Ch.III, Prop.6-11); 
 DeBrander (2007) links this tradition to the Stoics.

  The ethical imperative states that a given autopoietic system 
 should always try to increase the number of available choices,  
 diversify the possible ways and means it has to survive. 
  However, the necessary innovation processes should not be 
 shortsighted or improvident:  
  Time honored idioms teach us 
 not to be penny-wise and pound-foolish, 
 not to throw the baby out with the bathwater.  
  In order to increase its musical repertoire one has to learn new 
 songs, retaining however the melodies one already knows how to sing. 
  A viable path of evolution avoids unnecessary breaks in continuity  
 recycling already developed solutions, keeping them alive even 
 if as adjusted versions or in revised form.        
    
  Alongside all the innovations of Lavoisier's revolution, 
 there is also a tremendous effort to preserve the objects 
 of knowledge of the old (Stahlian) chemistry.  
  {\meuve Using this perspective, in Section 3 I analyze}   
 the ethical dimension of the new chemical nomenclature system 
devised by Lavoisier and Moerveau. 
  {\meuve Furthermore, I discuss the evolution}, during this period, 
 of the concept of chemical affinity,  focusing on its 
 algebraic additive structure and its capability (or lack thereof) 
 of working as the main organizing principle of chemical science.

\begin{figure}[h!]   
  \iflatextortf
   {} 
  \else
   \centerline{\includegraphics*[height=115mm,
      trim=19.0mm 30.5mm 18.0mm 19.0mm , clip]{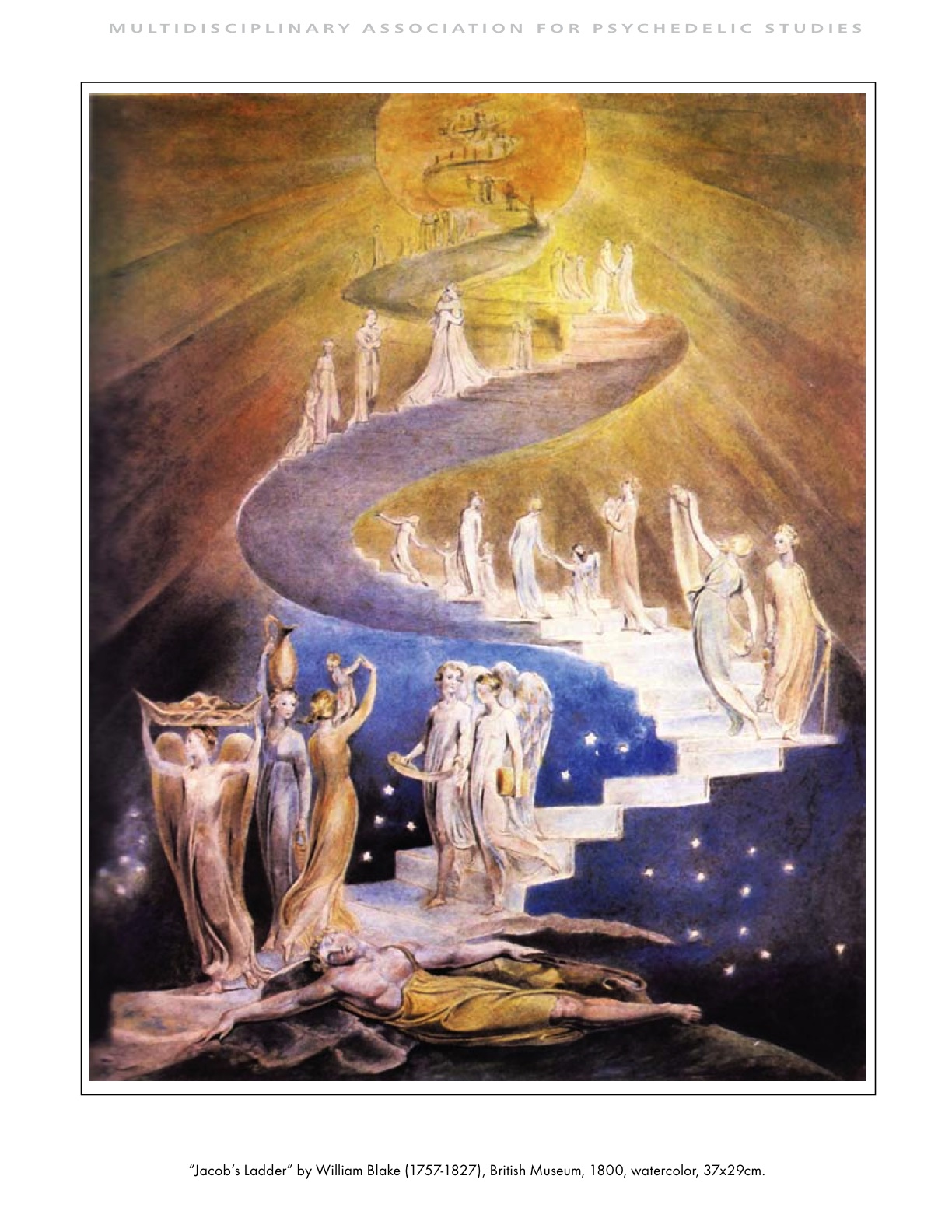}}     
   \centerline{\includegraphics*[width=150mm,  
           trim=0.7mm 0mm 1.4mm 2.2mm , clip]{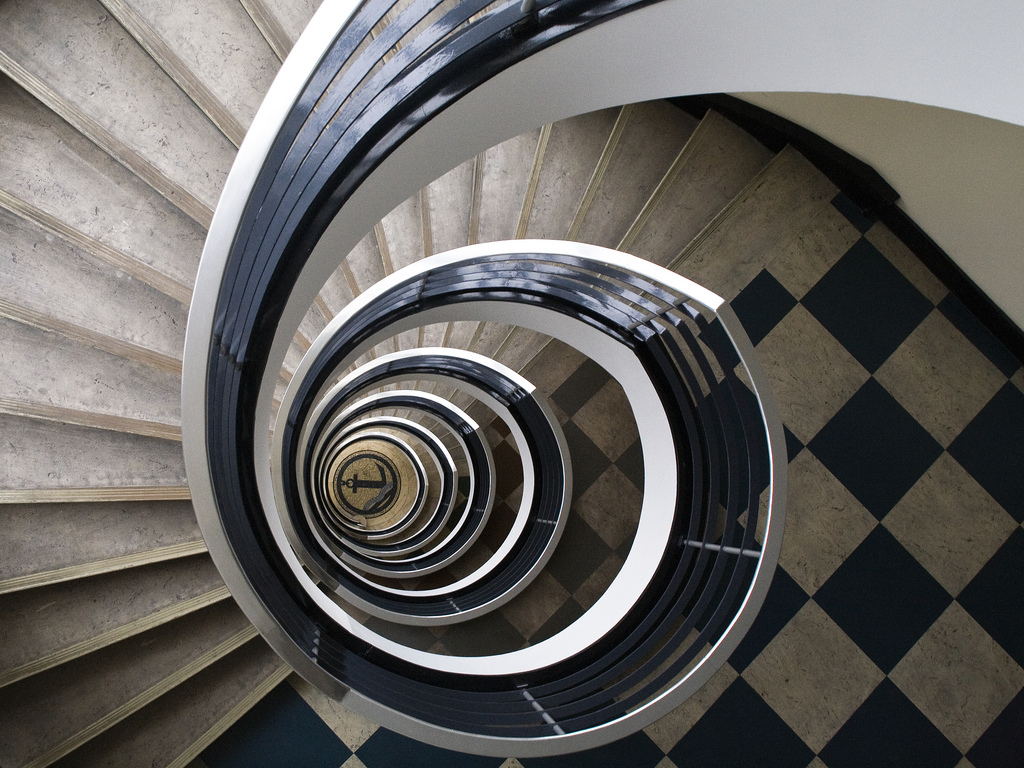}} 
  \fi 
   \vspace{0.5mm} 
   \centerline{Figures 2a: Jacob's Ladder by Willam's Blake;           
                     2b: Spiral staircase by Nils Eisfeld.} 
 \end{figure}


  Figure 2b displays a beautiful example of spiral staircase as it is 
 used in architecture.   
  The stairway from a given level to the next may represent the 
 theoretical and technological developments that allow the new 
 level to be reached.   
  Meanwhile, the building's floors linked by the staircase may be 
 considered as figurative analogs representing successive levels or 
 stages in the historical development of a scientific discipline. 
  At each one of such levels, the objects recognized by the scientific 
 discipline at that stage have to be organized in a coherent fashion. 
  This organization may be represented by the checkered or mosaic 
 pavement patterns at each floor of the building.

  A {\it scientific ontology} consists of a 
 {\meugr structured vocabulary for the} formal definition of the 
 collection of objects of knowledge of a given scientific discipline 
 and its organization, that is, the semantic relations that exist 
 between these objects. 
  Moreover, for the sake of continuity, it is important to compile 
 translation dictionaries or to establish valid (even if approximate) 
 correspondences relating objects of the present ontology of a scientific 
 discipline to objects of ontologies used in the immediate past or, 
 if possible, even to ontologies of much older times. 
  Employing our architectonic analogies, we must be able to map  
 positions in the current floor to corresponding positions in the 
 preceding or other previous floors. 
  {\meuve Such maps are known in computer science as 
	{\it ontology alignments}, these are briefly discussed
   Sections 2, 3 and 6.}  
 A good representation of successive ontologies of an evolving 
 science has them displayed in optimal alignment. 
  For straight stairways, a regular pitch or raising slope implies 
 successive shifts of relative horizontal position or changes in 
 orientation at the stair's landing points from floor to floor, 
 hence the preference for displaying Jacob's Ladder in spiral or 
 helical shape.  
    
  Metaphysics concerns causal explanations telling why things are 
 the way they do; these are the narratives, metaphors, and abstracts 
 symbolic statements used to build a system's understanding or intuition 
 about the world and the way it works, see Stern (2008b, Ch.4). 
   Von Foerster's (2003, p.284) alternative formulation of the 
 metaphysical (negative) imperative is:  
  ``For which we cannot show a cause or for which we do not
 have a reason, we do not wish to see.''  
  Section 4 illustrates the metaphysical (negative) imperative 
 in the (arguably somewhat retarded) development of spectroscopy.

  Science must evolve, acquiring valid knowledge over new objects, 
 or better (more objective) knowledge over its old domains. 
  However, as a consequence of the metaphysical imperative, 
 in order to be ready to see {\meuve some{\it thing}  }   
 we must already have some way of thinking about 
 {\it \meuve it}, we must already have some concepts able 
 to assimilate the structure of these ideas, and  
 we must have already crafted some terms in our language 
 that can be used to communicate these concepts.   
  Hence, the paradox: 
  How can we name or better describe what we do not yet see, 
 if we cannot see what we cannot yet explain?  

  The metaphysical imperative was formulated in negative form,
 and that is the way it ought to be.
  One cannot naively flip its orientation, stating it in
 positive form:  
 {\it \meuve Explain that which you want to see!} --- is just wishful thinking!     
  {\meugr  (Like in a M\"{o}bius band, in order to obtain a flip in orientation, 
 one has to travel all the way around it).}    
  In order to introduce new objects in a (upgraded) scientific 
 ontology, one has to reach a higher level of theoretical understanding, 
 developing new ideas and acquiring more powerful insights.   
  {\meugr Using once again the visual metaphor of Jacob’s Ladder,
 one has to climb up all the way around an helical stairway, reaching 
 the next floor of  the building, even if the building, the next floor, 
 or the stair itself,  is still under construction.}  
  Section 5 illustrates how such creative change is possible, 
 using, as historical examples, Lavoisier's early ideas of 
  {\meuve air-like} substances and Balmer's geometrical models.       
   In Section 6 {\meuve I present} my final remarks, and describe 
	some topics to be discussed in following articles.

 \subsection{Heavy Anchors and Fine Points}

   As in many representations of Jacob's Ladder, the images at 
  Figures 2a and 2b have the ladder's base firmly anchored 
  to the ground.   
   In our epistemological context, such heavy anchors are provided 
  by formal methods of Bayesian statistics developed with the 
  specific purpose of measuring the credibility of statements 
  declaring the existence of an eigen-solution --- technically,  
  computing an $e$-value, the epistemic value of a sharp 
  statistical hypothesis. 
   None of these formal methods will be directly used in the 
  present article. 
   However, the reader should be aware of their existence, for they 
  provide the nexus for application of the epistemological framework 
  under consideration to the practice of science and technology.  
   Without having these heavy methodological anchors readily 
  available and implemented as statistical tools for data analysis,  
  all the angels in William Blake's painting at Figure 2a  
  would remain in heaven, never reaching earthly realms or,  
  at least, not being able to reach empirical science's daily life.     
   The following sub-sections (1.1.1 -- 1.1.3) can be skipped on first 
	reading without loss of continuity; they contain some remarks on 
  statistical inference, observations on the interpretation of 
  eigen-solutions, and also some warnings on 
  historiographical methodology\footnote{  \meuve 
	In order to give firm foundations to the discussions in this paper, 
	there are many details about the chemistry of reaction networks that 
	must be carefully described. 
	 In particular, {\meuve I should} clearly define several alternative 
	conceptions of a reaction network {\meuve I am} using, and how 
	 they relate to each other. 
	 In order to do that, {\meuve I also} have to describe their 
	(sometimes implicit) algebraic representations, the 
	{\meuve compositional properties} of their 
	quantities of interest, like (algebraic) rules for handling stoichiometric 
	and affinity values, etc.  The interested reader can find this material in 
	 Stern  and Nakano (2014).}.

  \subsubsection{Bayesian Epistemic Value of a Sharp Hypothesis} 

   The essential properties of an eigen-solution, studied in 
  Section 2, dictate that these statements have the special 
  form of {\it sharp} or {\it precise} statistical hypotheses, 
  implying many important and non-trivial consequent requirements 
  for appropriate methods of statistical inference.        
   In the last 15 years, the Bayesian Statistics research group of  
  IME-USP, the Institute of Mathematics and Statistics of the 
  University of S\~{a}o Paulo,  has been developing a significance 
  measure known as the $e$-value --- the epistemic value of 
  a sharp or precise statistical hypotheses, $H$, 
  given the observational data, $X$; see Madruga (2001), Pereira 
  and Stern (1999) and Pereira et al. (2008).  
   At the same time, we have been exploring the {\meuve strong 
 algebraic structure of the compositional properties of  $e$-values, 
 that are characteristic of abstract belief calculi or 
 logical inferential systems},   
 see Borges and Stern (2007), Stern (2004, 2011a) and 
 Stern and Pereira (2014). 
   {\meuve These algebraic or logical properties further reflect  the 
	essential compositional properties of the eigen-solutions 
 the corresponding statistical hypotheses stand for}, 
 as discussed in Section 2.  
  Furthermore, {\meuve I have} been developing a specific version of 
 cognitive constructivism as a suitable epistemological counterpart 
 for this novel statistical theory, and vice-versa,  
 see Stern (2003, 2007a,b, 2008a,b, 2011a,b, 2012).

  The relationship between Bayesian $e$-values and the 
 epistemological framework of cognitive constructivism, 
 together with its fundamental metaphor of  
 {\it objects as tokens for eigen-solutions}     
 runs parallel to the relationship between 
 p-values of frequentist (classical) statistics and the 
 epistemological framework of Popperian falsificationism,
 together with its fundamental metaphor of  
 {\it the scientific tribunal}   
 where hypotheses are judged and proven (or not) to be false.  
  These associations, in turn, 
 run parallel to the relationship between   
 Bayes factors of decision-theoretic statistics and the 
 epistemological framework of  von Neumann-Morgenstern utility theory, 
 together with its fundamental metaphor of 
 {\it the scientific casino and rational betting systems.}  
  These parallel associations of inference systems and epistemological 
 frameworks are carefully compared  
 in Stern (2011a) and Stern and Pereira (2014).

 \subsubsection{Polysemy, Significance and Nature of Eigen-Forms}  
     
  \noindent    
  {\bf 1.1.2.1 \  Metaphorical meaning vs. formal structure of eigen-forms:} 
  
  Each one of the aforementioned statistical frameworks, 
 or just their corresponding fundamental metaphors, can be used as 
 ways of thinking about and understanding the world we live, 
 far beyond of hard empirical science.   
       
  Game theory, originally developed by Morgenstern and von 
 Neumann (1947) in the context of economy, was soon adapted for 
 a new role in foundations of Bayesian statistics, 
 see Blackwell and Girshick (1954), Dubins and Savage (1965)  
 and  Finetti (1972, 1974). 
  Afterwards, the scope of game theory was extended to ethology, 
 psychology, epistemology, and even foundations of mathematics,  
 see Jech (1973, Sec.12.3 -- The Axiom of Determinateness) 
 and Shafer and Vovk (2001, Sec.4.6).   
  However, as the scope of possible applications of game theory 
 expands, it is unavoidable that the word {\it game} becomes 
 increasingly polysemic. 
  Nevertheless, reasonable applications of game theory share 
 strong formal properties captured by well defined mathematical 
 structures, preserving a methodological integrity in the field 
 and  allowing  coherent interpretations in different applications. 
  This issue is strongly connected to the theme of ontology  
 alignment, that is commented throughout the text, and to von 
 Foerster's organic imperative, briefly commented on in Section 6.        
 
  Moreover, examining the literature of utility / game / 
 decision theory, we can attest undeniable positive-feedback, 
 reinforcement and synergy effects between 
 (1) the development of mathematical formulation of these theories; 
 (2) the development of the corresponding epistemological frameworks;   
 (3) the development of Bayes Factors and other analytical tools based 
     on decision-theoretic Bayesian Statistics; and   
 (4) a coherent generalization of the underlying game or 
     gambling metaphors.

  {\meuve I believe} that similar synergy effects are possible between 
 (1) the use of an already vast and ever expanding literature on 
     mathematical models for or based on functional eigen-solutions; 
 (2) the development of the epistemological framework of 
     cognitive constructivism;   
 (3) the development of analytical tools based on Bayesian measures of 
     epistemic significance for sharp statistical hypotheses and their 
     {\meuve compositional properties (logical rules)}; and  
 (4) a coherent generalization of the underlying metaphor of 
      {\it objects as eigen-solutions.}   
   One of the main goals of our research group is to study and 
  foster this synergy, hence our emphasis on formal and quantitative 
  analyses for characterizing empirical eigen-solutions.  
   

  \noindent    
  {\bf 1.1.2.2 \ Constructive vs. objective nature of eigen-solutions:} 

   In the epistemological framework of cognitive constructivism, 
  eigen-solutions emerge from the recursive interactions 
  of an autopoietic system with its environment. 
   Hence, any object known by an organism or abstract autopoietic 
  system has an undeniable constructive character. 
   Better said, as products of an interaction process, objects 
  must depend on the interacting parts: 
  The observer and the observed; The actor and what it acts upon; 
  {\meuve Some{\it thing}} out there and the one manipulating 
	{\it \meuve it}.   
   This delicate theme concerns the {\it self-reference paradox},  
  studied at Stern (2007b, 2008b Sec.6.4), 
  see also  Brier (2001, 2005) and Rasch (2000, Ch.3,4). 

   In our framework, {\it objectivity} refers to the truth-value 
  of the known object, that is, the degree in which it exhibits the 
  essential properties of the underlying eigen-solution, namely, 
  precision, stability, separability and composability.  
   Moreover, in scientific applications, such (statistical) 
  {\it significance measures} are essential for the discovery, 
  calibration, and standardization of the interactive processes 
  in which an object emerges. 
   It is through this fine-tuning that the entities in a scientific 
  ontology are ultimately defined.   
   The historical examples presented in this paper should help 
  to make it clear how this process takes place.   
  
    The constructive character of eigen-solutions can be illustrated 
  by several examples in Cerezetti (2013).   
   The author develops algorithms for automatic detection of 
  {\it arbitrage opportunities} --- events recognized  
  by specific failures in equilibrium conditions 
  that characterize {\it efficient markets}.  
   These algorithms have been used and tested at the 
   BM\&F-Bovespa  derivative markets.    
   The (statistical) detection of an arbitrage opportunity  
  can, in turn, have two interesting applications: 
   On the one hand, it can be used for speculative trading;   
   On the other hand, it can be used by market regulators, 
  preventing fraudulent trading, activating circuit breakers 
  or taking other pertinent exception handling procedures. 
   In either of these two roles, speculator or regulator, 
  the agents using such algorithms act as fingers of the 
  {\it invisible hand}, effectively building the efficient markets 
  in which eigen-values known as {\it prices} can emerge having the 
  desired meaning and operational properties. 
   
   Players and regulators of finacial markets,  small 
  communities issuing complementary currencies, as in Figure 1, 
  and birds singing and synchronizing their songs, as in Figure 4,  
  are all actively {\it \meuve trying} to build stable eigen-forms.  
   Alas, using variational principles, we can give teleological 
  explanations even to the movements of infinitesimal particles,  
  see Stern (2011b).    
   Nevertheless, the level of  conscience or self-awareness    
  of the role played by {\it agents} participating in each one 
  of these construction projects / processes varies greatly.             
   Symbolic meaning, cognition and communication of eigen-forms at 
  different levels of complex hierarchical structures are topics  
  examined in Brier (2001, 2005).

 \subsubsection{Historiographic Remarks}

 Before ending this section, it is only fair to post a 
 {\it \meuve caveat emptor} about the illustrative use made in this article 
of some examples in the history of science.   
 If establishing a single fact may be a challenging task, interpreting 
a dynamical processes is inevitably a lot harder, especially so at 
times where concepts are in rapid flux and language undergoes great
transformations.  
 Nevertheless, the best examples to illustrate the epistemological 
ideas presented in this paper are offered exactly by processes 
taking place in such {\meuve eyes of the storm}. 
  {\meuve I have} chosen examples from the turn of the 
	{\meuve 18th to the 19th century.}   
 Choosing more recent alternatives could, perhaps, alleviate some  
historiographical difficulties, but would also impose heavier 
overheads of mathematical formalism and specialized scientific 
knowledge.   

  Furthermore, a word of caution must be given about the references in
 the literature of history of science listed in the text. 
  The cited authors often come from very different backgrounds,
 use incongruent methods of analysis, and may produce 
 divergent general conclusions.  
  Hence, one should not expect the aforementioned references 
 to provide an overall harmonious and coherent perspective.  
  Nevertheless, in each of such references {\meuve I could} find a 
 specific point of view or an interesting comment that 
 {\meuve (in my opinion)} is somehow supportive or pertinent to the 
 epistemological ideas being developed in this article. 
  Furthermore, the fact that two or more authors, coming from 
 diverse schools of thought and different academic traditions, 
 can agree to distinguish certain characteristic    
 in a specific thread of historical process  
 or indicate a distinctive aspect of a given event, 
 may be taken as corroborative evidence in favor of 
 these common factors or shared interpretations. 
  Finally, {\meuve I hope} that this article motivates further and 
 far more technical inquires in history of science. 


 \section{Objective Aesthetics and Eigen-Solutions} 

  Eigen-solution is the key concept of the objective version of 
 cognitive constructivism used in this article. 
  Eigen-solutions emerge as operational fixed-points or equilibria,  
 as invariants for a system interacting with its environment. 
 {\meuve 
  A fundamental insight for the constructivist framework is expressed
	in my re-interpretation of von Foerster’s well-known metaphor as: 
	Objects are tokens for eigen-solutions. }   
  In other words, objects, and the names we use to call them, 
 stand for and point at the eigen-solutions that emerge in our  
 eigen-behaviors, that is, in the stable recurrent interactions 
 we have in our environment.   
  Figure 3 displays some postage stamps; these are tokens 
 (that can be exchanged) for a specific and well defined postal 
 service. The themata depicted in these stamps are also related 
 to matters discussed in this article.

 \begin{figure}[tb]  
  \iflatextortf
   {} 
  \else  
  \centerline{
  \includegraphics[height=1.2in, width=1.5in]{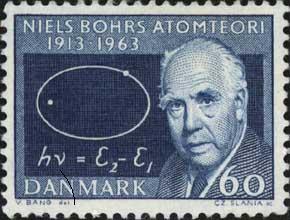} 
  \includegraphics[height=1.2in, width=2.3in]{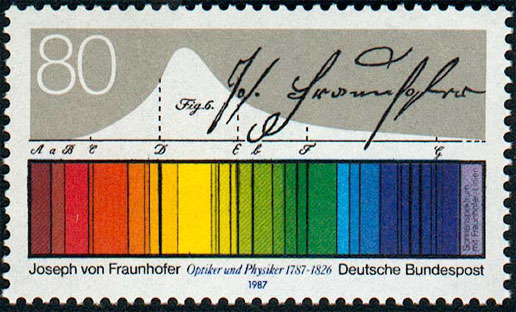} 
  \includegraphics[height=1.2in, width=1.3in, 
   trim=0.2in 0.2in 0.2in 0.2in, clip]{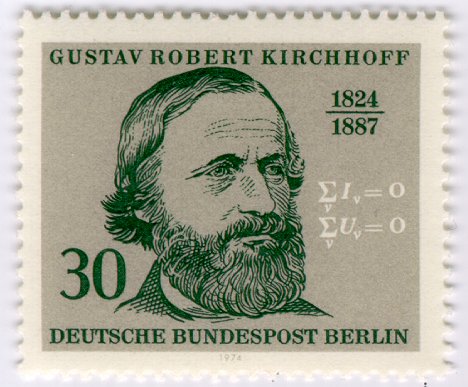}   
  \includegraphics[height=1.2in, width=1.1in]{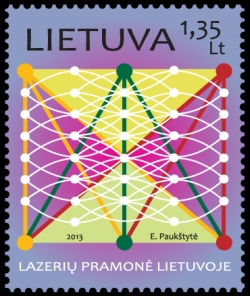} }
 \fi 
 \vspace{0.5mm} 

 \centerline{Figures 3abcd: Scientific aesthetics: Nice colors, 
             sharp images, and beautiful tokens.} 
 \end{figure}

  When a musician plucks a string of his or her guitar, 
 he or she produces very definite musical notes. 
  In modern physical language, these notes emerge as eigen-solutions 
 of the (small amplitude) wave-equation for the (homogeneous) string. 
  For mathematical, graphical and artistic representations of these 
 eigen-solutions, the fundamental and higher harmonic standing-waves, 
 see Stern (2008a, p.50-53) and Figure 3d.   
  Nevertheless, humans have been playing 
 {\meugr guitars, harps and similar instruments long before anyone 
 knew how to solve a differential equation, see Figure 4a}.  
  That is, humans have perceived the existence of  {\it objects}  
 they called musical notes, and used them as the key elements for  
 playing or composing music, long before they had the abstract 
 mathematical notion of a functional eigen-solution. 
      
  {\meugr Figure 4b} presents a short song using modern 
	musical notation. 
  However, this is not a human song, but an archetypal form   
  from the Plain-tailed Wren's four-part chorus, see Mann (2006). 
  Speciation is a biological evolutionary process in which such  
 eigen-forms emerge as stable organizational patterns. 
  According to some schools of neo-Jungian psychiatry, 
 evolutionary eigen-behaviors known as {\it archetypes}  
 play a primary role in human psychology,     
  see Stevens (1999, 2003) and Stevens and Price (2000).   
  {\meugr Figure 4c} displays the beautiful plumage of the Scarlet 
 Tanager (Ramphocelus bresilius) in a painting by John James Audubon. 
  The Ti\^{e} Sangue is in Brazil a symbol of freedom, for the 
 vivid colors that a healthy bird displays in the wild quickly fade 
 away when it is held in captivity.
  This bird is also displayed at the back cover of 
 Ber van Perlo's  {\it A Field Guide to the Birds of Brazil}, 
 a catalog of archetypal forms used by bird watchers  
 for taxonomic identification.

  Musical notes are the {\it atoms} of Western music:  
  In the most basic form, they constitute a {\it discrete} set of  
 (fundamental) frequencies available for musical performance.  
  The (relative) frequency of each of these notes is exactly defined. 
  {\it Precision}, that is, having an {\it exact} or {\it sharp} 
 definition, is the first essential property of eigen-solutions.  
  The guitar is a musical instrument that is carefully designed in 
 order to make it easy to produce these musical notes. 
  Small variations in the way the player plucks a string will not 
 substantially change (the fundamental frequency of) a musical note,  
 neither will reasonable variations in the atmospheric conditions, 
 and so forth.  
  This property is called {\it stability}, the second essential 
 property of eigen-solutions. 
  It is also possible to produce singular tones separately, 
 or to  produce simultaneously several notes without distorting 
 each of the singular tones.  
  Eigen-solutions are {\it separable} and {\it composable}.  
  These essential properties allow the artist to use musical notes 
 as building blocks in the construction of harmonious chords and 
 beautiful melodies.


 \begin{figure}[bt]  
  \iflatextortf
   {} 
  \else  
  \centerline{
  \includegraphics*[height=1.2in, angle=0, 
    trim=0.0in 0.0in 0.0in 0.0in, clip]{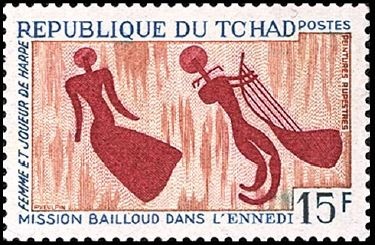} 
  \mbox{} \ \mbox{} 
  \includegraphics*[height=1.1in]{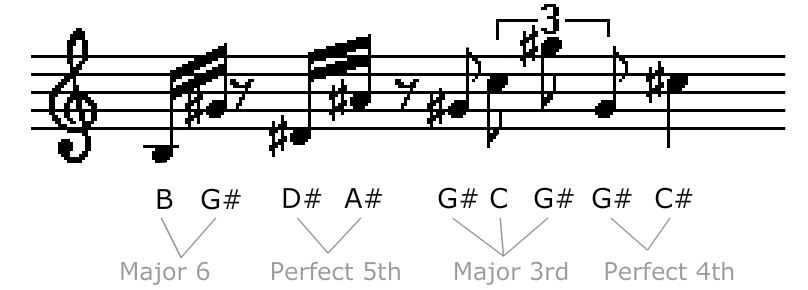} 
  \mbox{} \ \mbox{} 
  \includegraphics*[height=1.2in, angle=0, 
    trim=0in 0in 0.93in 0.42in, clip]{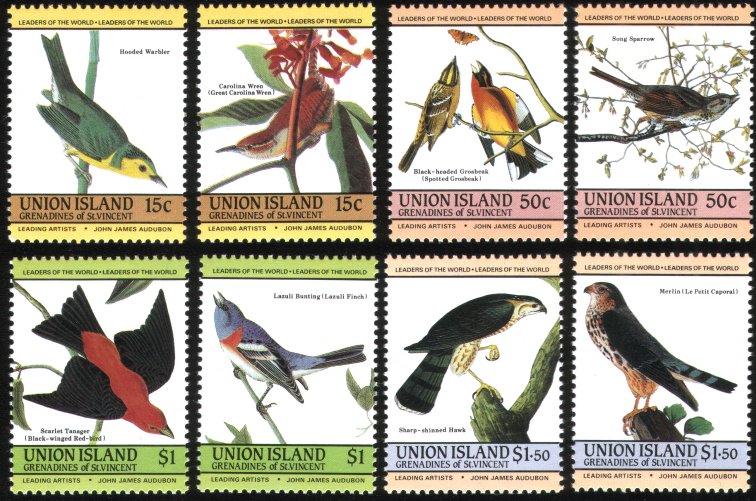} 
   }
 \fi 
 \vspace{0.5mm} 

 \centerline{Figures 4abc: 
    Archetypal forms in music and biological speciation.} 
 \end{figure}

 {\meuve To Summarize:} The four essential properties of 
 eigen-solutions are to be discrete (or sharp, exact or precise), 
 stable, separable and composable.     
  In the constructivist framework, these four essential 
 properties define the possible forms of perception of an 
 eigen-solution, that is, the aesthetical attributes of the 
 corresponding scientific object, 
 the objectivity of the same science and, perhaps, 
  also the nicety of related experiments 
  or the beauty of pertinent demonstrations;     
 see Foerster (2003c, p.266), Golinski (1995), 
 Segal (2001, p.145, 127-128), Stern (2007a, 2008a, 20011a), 
 and Szab\'{o} (1978, p.240).
 
  Eigen-solutions can also be tagged or labeled by words, and these
 words can be articulated in language. 
  Of course, the articulation rules defined for a given language, 
 its grammar and semantics, only make the language useful if they 
 somehow correspond to the composition rules for the objects the 
 words stand for. 
  Ontologies are carefully controlled languages used in the practice 
 of science\footnote{  \meuve 
	  Nowadays, the development of some areas of empirical science 
would be impossible without powerful computational tools for 
ontology management.
  For example, current research in genomics requires well structured 
and carefully controlled vocabularies for identification of biological 
genes, and subsequent annotation concerning: 
sequencing alternatives;  proteomic translation; biochemical, 
cellular and extra-cellular function; and so forth. 
 Moreover, a good ontology management framework should provide 
ontology alignment and other statistical analysis tools capable of 
tracing evolutionary relations, following biochemical pathways, 
identifying similarities in function across biological species, 
and searching for other meaningful correspondences.  
 Furthermore, such a computational framework should 
enable the integration of large, distributed and diverse databases, 
facilitating collaborative work of independent research groups.}. 
  They are developed as tools for scientific communication.
  According to the constructivist perspective, the key words in 
 scientific ontologies are {\meuve labels for} eigen-solutions. 
  This is the constructivist approach to the classic problems of 
 external symbol grounding and alignment of scientific ontologies, 
 as briefly discussed in Section 6.

  The importance of working with eigen-solutions is illustrated by 
 the elegant origami example, presented by Richard Dawkins in 
 Blackmore (1999, p.x-xii). 
  Dawkins compares two instances of the Chinese whispers game.   
  In this game, an object is shown to the first child in a line of 
 children, that must copy the object the best he can, and then show 
 the copy he made to the second child in the line, and so on. 
  The fidelity of the copy mechanism can be estimated by the 
 difference between the original object shown to the first child,  
 and the copy produced by the last child in the line.  
  Dawkins than compares the results obtained when copying two very 
 distinct objects: 
  A freehand drawing and an origami, both representing a Chinese junk.  
  The result is easy to foresee:   
  Except for the rare occurrence of a spurious mutation, the successive 
 copies of the origami are all identical, that is, most of the time, 
 despite small differences in craftsmanship, all copies follow 
 the exact same folding scheme. 
  Meanwhile, the final copy of the freehand drawing has little  
 resemblance to its original.  
  These results are similar to the progressive degradation 
 afflicting successive copies of an analogous tape recording, 
 in contrast to the high-fidelity copying process for digital discs.   

  \begin{figure}[tb] 
  \iflatextortf
   {} 
  \else  
   \centerline{  
    \begin{tabular}{c}
    \includegraphics[height=2.1in, width=2.5in]{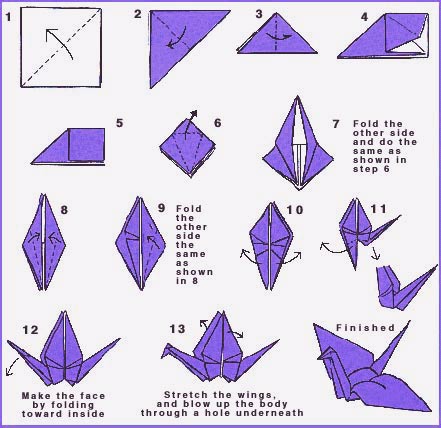} 
    \end{tabular} 
     \mbox{} 
    \begin{tabular}{c} 
      \includegraphics[height=1.0in, width=1.5in]{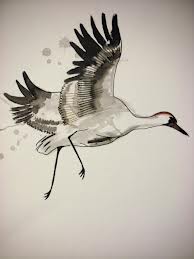} \\ 
      \includegraphics[height=1.0in, width=1.5in]{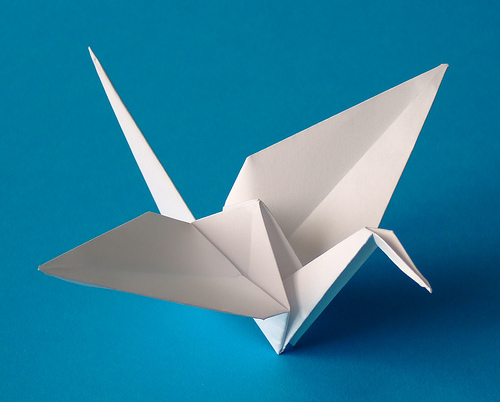} 
    \end{tabular}  
  } 
  \centerline{
    \includegraphics[height=1.8in, width=2.1in]{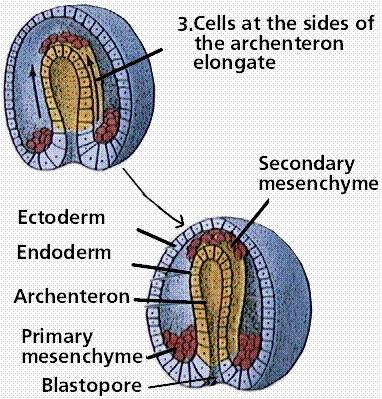}
    \mbox{} 
    \includegraphics[height=1.8in, width=2.0in]{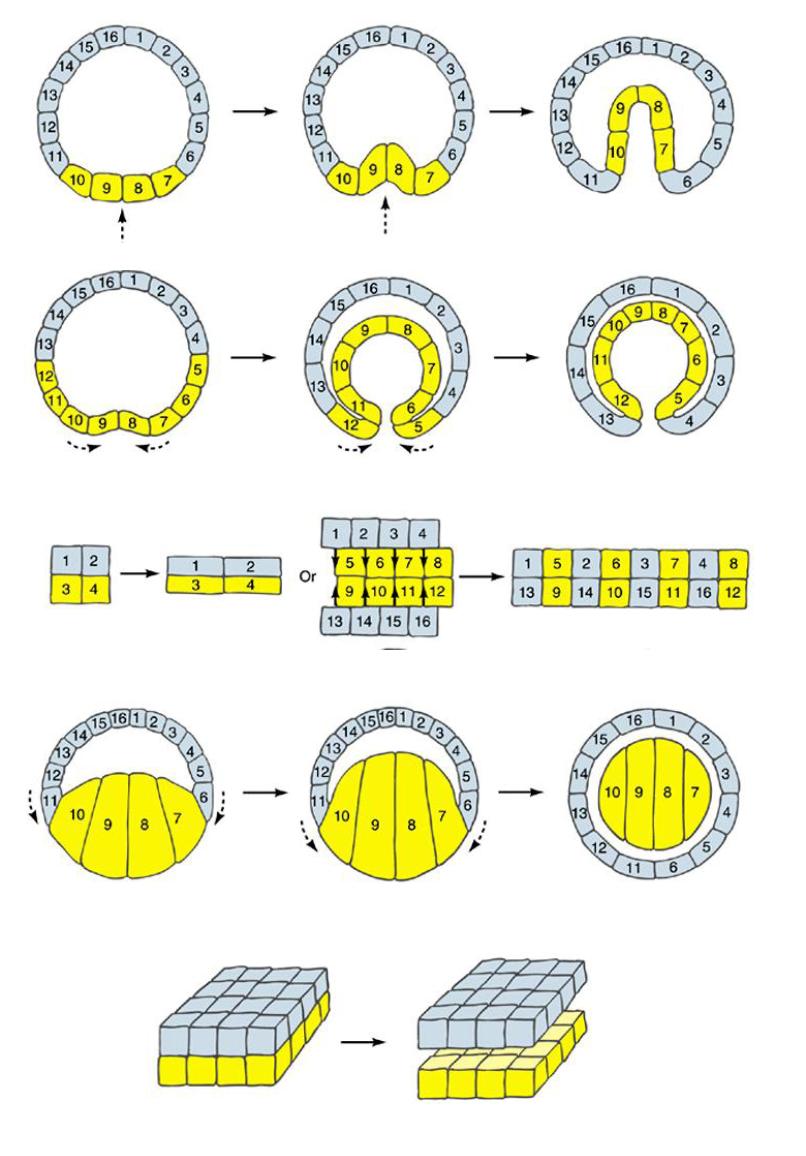}     
   }
  \fi 
  \vspace{0.5mm} 
  \centerline{Figures 5abcde: Geometrical foldings, origamic and organic.} 
  \end{figure}

    An origami is produced by a successive folding process. 
   The small number of basic folds used in the traditional art of 
  origami have the same essential characteristics of eigen-solutions: 
   They are defined as exact geometric operations, that can be 
  produced with great precision and stability. 
   Furthermore, many successive folds can be easily superposed  
  in order to produce complex compositions.  
   Even larger and more intricate structures can be assembled by 
  modular composition of individual origami objects.  
   Figure 5a shows the instruction sheet for the {\it tsuru},   
  the classical origami for a crane. 
   Real (biological) cranes are self-assembled by organic 
  morphogenesis.  
   Figure 5e depicts tissue foldings used in this process: invagination, 
  involution, convergent extension, epiboly and delamination.   
   Of course, analyzing eigen-solutions of complex biological processes 
  demands  sophisticated tools, like stochastic processes, statistical 
  models, fractal geometry, catastrophe theory, and so forth.

  \subsection{Blowpipe and Spectral Analysis} 
   
  {\meuve Let me} now present some examples of eigen-solutions in 
 the field of chemistry, comparing two techniques used to identify 
 chemical substances, namely, blowpipe and spectral analysis.  

   A typical blowpipe consists of a small curved brass tube  
 having at its ends a mouthpiece and a fine nozzle. 
   Blowing through this tool, an experienced operator can direct 
 a jet of air through a gas burner, obtaining a flame with 
 special characteristics (oxidizing or reducing) and high 
 temperatures at specific spots.  
  Taking small samples of chemical substances into this flame 
 produces light with characteristic colors, as displayed in 
 Figure 6a, that can be used to identify the presence of known 
 chemical elements. 
  
  In 1751, Axel Fredrik Cronstedt used blowpipe analysis to 
 identify a new chemical element, the metal nickel. 
  Carl Wilhelm Scheele and his coworkers used blowpipe analysis to 
 identify the new elements manganese, 1774, molybdenum, 1781, and 
 tungsten, 1783.  
  From that point on, this tool contributed for the discovery of 
 several new chemical elements, see Jensen (1986). 
  Nevertheless, standard blowpipe analysis also has some drawbacks. 
  Among them, is the somewhat subjective nature of the specific 
 color tones and hues that can be seen at a flame, the recognition 
 of which requires extensive training of the analyst and, even then,
 is a task prone to error. 
 
 \begin{figure}[tb] 
  \iflatextortf
   {} 
  \else 
   \centerline{  
    \begin{tabular}{c} 
   \includegraphics[height=1.4in, width=2.2in]{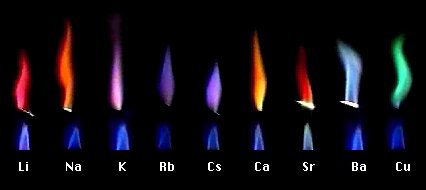}  
	\end{tabular}  
     \mbox{} 
    \begin{tabular}{c}             
    \includegraphics[height=1.0in, angle=0, 
                 trim=0.0in 0.0in 0.0in 0.0in, clip]{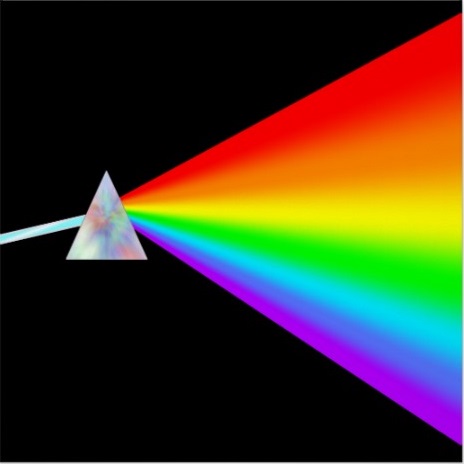} 
    \end{tabular} 
     \mbox{} 
    \begin{tabular}{c}
    \includegraphics[height=1.4in, width=2.4in]{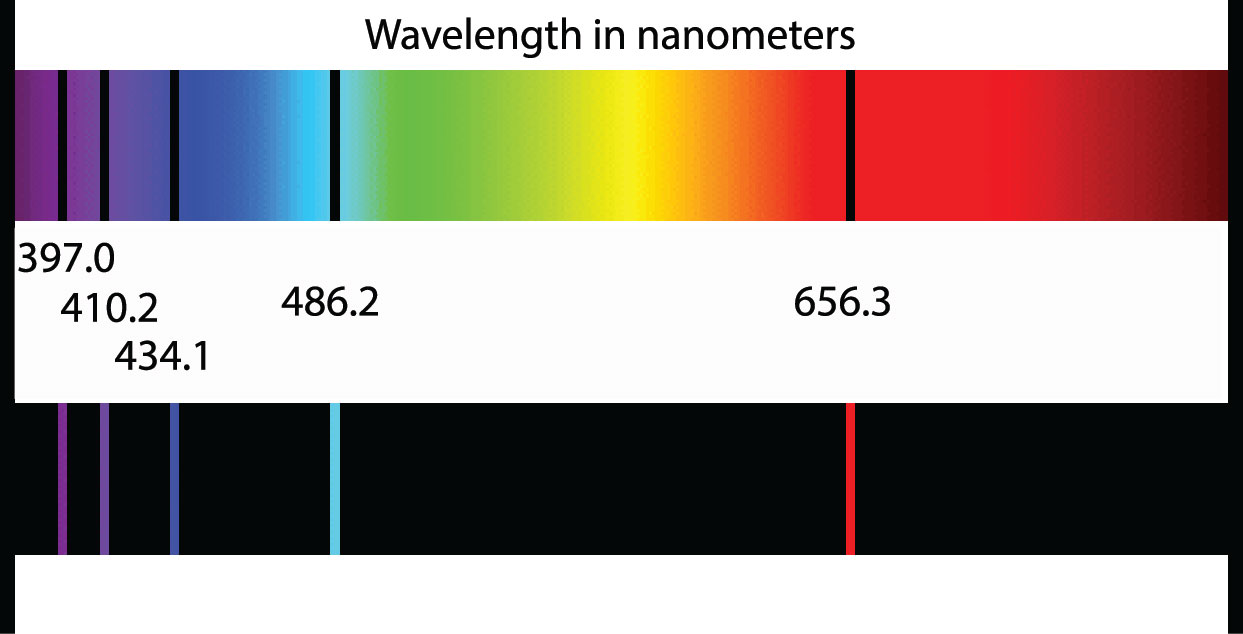} 
    \end{tabular} 
   } 
 \fi 
 \centerline{ \meuve Figures 6a: Blowpipe colors; 
             6b: Dispersion by refracting prism; 6c: Spectral lines.} 
 \end{figure}

  Spectroscopy is a technique that overcomes many shortcomings 
 of blowpipe analysis. 
  Since ancient times it is known that a prism can decompose light 
 into its constituent colors, making it possible to analyze the 
 resulting spectrum. 
  Isaac Newton describes several spectroscopic experiments in his 
 book Optics of 1704. 
  In 1785 David Rittenhouse was able to manufacture diffraction 
 gratings with approximately 100 lines per inch, a better alternative 
 for spectroscopy than prisms.   
  In 1802  William Hyde Wollaston reported dark lines in the 
 spectrum of solar light. 
  Between 1814 and 1821 Joseph von Fraunhofer build the first 
 reliable spectroscopes, first using prisms, and later replacing  
 them by diffraction gratings.  
  Fraunhofer analyzed the spectra of natural sun light and also 
 alternative sources like a variety of combustion flames, 
 observing with great accuracy absorption and emission spectra. 
  Emission spectra are composed by a discrete set of bright lines, 
 each one defined by a precise position that corresponds, in 
 contemporary physical theory, to an exact frequency or wave-length.  
  Absorption spectra exhibit a complementary picture, that is, an 
 continuous color composition except for a finite set of dark lines. 
  Figure 3b shows Fraunhofer's solar absorption spectrum and 
  Figure 6c shows the emission and absorption lines of hydrogen.    
   
  In 1859, Gustav Robert Kirchhoff, see Figure 3c,   
 and Robert Wilhelm Eberhard Bunsen realized that spectroscopy analysis 
 could be a highly efficient method for analytical chemistry, 
 starting a revolution in the field, see  Kirchhoff (1860a) and 
 Kirchhoff and Bunsen (1860).  
  Many historians have wondered why this revolution took so long, 
 because, as one can infer from the aforementioned chronologies,   
 see Figure 8, all necessary the technology was available for quite 
 some time.     
   {\meuve I will} analyse this question in great detail in section 4. 
  For now, {\meuve I just want} to call the reader's attention for the 
 possibility of a parallel correspondence between, on the one hand, the 
 blowpipe and spectral analysis methods described in this section and, 
 on the other hand, the freehand drawing and the origami instances  
 of the Chinese whispers game described in the last section.

 In 1885, Johann Jakob Balmer was the first to achieve an explanation 
for the relative position of the emission lines of hydrogen, 
the simplest of all atoms.  
 The frequencies, $\nu$, or wavelengths, $\lambda$, of these
spectral lines are related by an integer algebraic expression:  
 \iflatextortf  
  \mbox{} \ \ \ \ \ \ \ \ \ \ \ \ \ \ \
     $\nu(n,m)/c  \ =  \ 1/\lambda(n,m)  \ =  \
     R \left( 1/n^2 \ - \ 1/m^2 \right)\ ,$ \ \mbox{} \\

  \noindent 
 \else  
  \[
    \frac{\nu_{n,m}}{c} = \frac{1}{\lambda_{n,m}} =
    R \left(
    \frac{1}{n^2} -\frac{1}{m^2} \right) \ ,
  \]
 \fi 
 where $R=1.0973731568525(73) E7 \ m^{-1}$ is Rydberg's constant 
 (this notation includes a standard deviation for the last 
 significant digits, 525$\pm$73). 
  In Balmer's formula, distinct combinations of integer numbers, 
 {\meuve $0<n<m$}, give wavelengths of distinct spectral lines, 
 see Balmer (1885, 1897), Banet (1966, 1970) and Enge (1972).     
  Further comments on Balmer's intuition for arriving at his 
 formula are given in Section 5.

  Many times, when great scientists face fundamental eigen-solutions, 
 they are taken by a marvelous sense of wonder or find themselves  
 in a state of ecstatic admiration, as expressed by Balmer in the 
 following quotations.  
 
  \begin{quotation}
 {\myqt Extending the calculations based on the last estimated constants
 to the following (spectral) lines, resulted in an average deviation
 from the measured wave-lengths of only about 1/4 of a unit.
    Obtaining such a closely correct result
 in the first trial of this formula -
 using only round integer numbers for the first and second constants,
 was for me a great surprise, that strengthened to the highest degree
 my conviction that this formula is the most adequate expression
 for a physical truth.}
 Balmer. (1897, p.383)
 \end{quotation}

  \begin{quotation}
 {\myqt The final impression, involuntarily imposed on our spirit by
 contemplating such a basic and fundamental relationship,
 is the existence of an inexhaustible wisdom established in nature
 that fulfills its function with infallible certainty,
 a wisdom that a thinking spirit can only perceive in incompleteness,
 following it with humbling and arduous efforts.}
 Balmer. (1897, p.391). 
 \end{quotation}

  Perhaps, these strong feelings could be taken as a helpful hint,  
 suggesting that the constructivist approach to objectivity provides 
 sensible intuitions.  
  Maybe, these strong subjective impressions reveal an a priori 
 conception indicating that the constructivist notion of reality 
 offers powerful insights.

  \subsection{Stoichiometry, Conservation Laws 
              and Invariant Elements}

  The example of scientific eigen-solutions seen in the last 
 sub-section is expressed by Balmer's formula, a mathematical 
 equation. 
  That is exactly the case for the most fundamental scientific 
 statements. Well known examples of deterministic statements are  
 Newton or Einstein's laws of gravitation or Maxwell's equations for 
 electrodynamics; good examples of probabilistic statements can be   
 found in statistical physics, stochastic equations of population 
 genetics or in formulas for state transition probabilities in 
 quantum mechanics.   
  In all these examples, the equality sign of the law, formula or 
 equation, expresses the first essential quality of an 
 eigen-solution, namely, precision.

  Even considering that any actual experiment aiming to verify 
 a scientific statement expressed as a mathematical equation 
 has its design flaws and is plagued by a variety of 
 operational imprecisions and measurement errors, there is an  
 underlying equality relation the experiment aims to access. 
  Modern technology is living proof that, to a great extent, 
 science has been very successful at this task. 
  As a trivial example, the Intel CPU powering the gadget 
 {\meuve I am} using  to write this article is a complex composition 
 of about 200 million transistors, designed to implement sophisticated 
 Boolean logic algorithms. 
  Furthermore, it operates at a clock rate over 3 GHz, so that 
 all and every single one of these transistors must operate in 
 time synchrony of less than half a part per billion 
 {\meugr of a human heart-beat!}     

  In this sub-section, {\meuve I make} some remarks about the 
	perceived value in Lavoisier's work of von Foerster's essential 
 characteristics of eigen-solutions, starting with precision.  
  {\meuve I argue} that the new standard of precision introduced by 
 Lavoisier was an important factor in the public (and his own) 
 perception of these changes  as a revolution that brought 
 chemistry into science, see also  
 Donovan (1988, 1990, p.270-272) and Perrin (1990, p.260).

  In 1774, the {\it  Acad\'{e}mie Royale des Sciences} invited 
 Lavoisier to write an anonymous review of his own work, the  
 {\it Opuscules Physiques et Chimiques.} 
  According to Perrin (1984), at that time, such an invitation 
 was considered acceptable practice. 
  Moreover, this self-referential ``anonymous''   review give us a 
 rare opportunity to get a sincere assessment of what the author 
 most appreciated in his own work.  
  The next two quotations give the first and last paragraphs of 
 this precious review.    
  The third quotation is also from Lavoisier; it is not anonymous 
 but it is meta-theoretical, in the sense that it brings the author's 
 opinion about the goals, means and methods of general science.

  \begin{quotation} 
 {\myqt  Mr. Lavoisier has published, at the beginning of 1774, a book 
entitled Physical and Chemical Essays. 
 The subject of this work was to examine the nature and properties of
these air-like fluids that flow out from or combine with the body. 
 These fluids have been hardly noticed by scientists until recently,
when they become, for the last few years, one of the main objects of
their research.} 
  Lavoisier (1774, p.89).   
 \end{quotation}

 \begin{quotation} 
 {\myqt Those are the main experiments contained in the work of 
 Mr. Lavoisier: he applies to chemistry, not only the equipment 
and the methods of experimental physics, but also the spirit
of calculation and precision  that characterizes this science.
  The union that appears to be in the making between these two branches
of knowledge will lead to a brilliant era of progress for both of them,
and Mr. Lavoisier is one of those that most contributed for this 
desired reunion we so long awaited.}  
  Lavoisier (1774, p.96).    
  \end{quotation}

  \begin{quotation} 
 {\myqt  I have already observed in preceding memories  
  that the way of reasoning is the same for all sciences; 
  that chemists, like geometers, can only proceed from 
  what is known to the unknown by true mathematical analysis,
  and that all reasoning in matters of science
  implicitly contain true equations.}   
 Lavoisier. (1788, p.777-778). 
  \end{quotation}

  The value Lavoisier gives for the precision of his experiments 
 takes him (perhaps unconsciously) to even exaggerate the exactitude 
 of his work, for example, showing calculated results with 
 more than the appropriate number of significant digits. 
  On the one hand, we must recognize that a good error analysis 
 theory was not available at that time.       
  On the other hand, Lavoisier clearly makes a rhetorical use of 
 this exaggeration, for it is clear to him that the more exact 
 are the results that he can show, the more scientific will be 
 the perception of his new chemistry. 
  Lavoisier's  friends and adversaries alike were well aware of the 
 potential and also of the liabilities of precision as a rhetorical 
 device, see Golinski (1995) and Levere (2001, p.93).    
  Nevertheless, according to the constructivist perspective, honest 
 reports of high-precision experiments is exactly what one should      
 expect from good (objective) science.

  According to the constructivist perspective, not only precision, 
 but also stability, separability and {\meuve compositionality} are 
 essential characteristics of good scientific objects.  
  Explicitly seeking these properties, Lavoisier suggests a 
 new list of chemical elements. 
  Moreover, Lavoisier gives this list a provisional status, 
 because more powerful analytical methods could always be 
 developed, showing how to reduce 
  {\meuve ``substances that had so far resisted analysis 
  into simpler  building blocks''}, see Levere (2001, p.69). 
  Furthermore, Lavoisier is careful not to tie his operational 
 definition of chemical element to any other unnecessary 
 hypothesis about their ``true nature''.     
  The following quotation should make these points clear.

 \begin{quotation} 
 {\myqt  If, by the term elements, we mean to express those simple and
indivisible atoms of which matter is composed, it is extremely probable
that we know nothing about them; but, if we apply the term elements, or
principles of bodies, to express our idea of the last point which
analysis is capable of reaching, we must admit, as elements, all the
substances into which we are capable, by any means, to reduce bodies by
decomposition.} 
 Lavoisier (1789,v.1, p.xvii), as quoted in Levere (2001, p.80).  
 \end{quotation}  

  {\meugr Jeremias Benjamin Richter coined the term {\it Stoichiometry} 
 to designate chemical balance equations giving the exact proportions 
 in which the involved substances interact.}  
 
 \begin{quotation}
 {\myqt  Because the mathematical part of chemistry is constituted mainly 
 by such bodies, which are indivisible substances or elements, and
 because  this science discusses the volume proportions between them,
 I could not find a shorter and more appropriate name for it than the
 word St\"{o}chyometria, of the Greek
 $\sigma \tau o \iota \chi \epsilon \iota o \nu$,
 which means something like indivisible, and the word
 $\mu \epsilon \tau \rho \epsilon \iota \nu$,
 which means the search for volume proportions.} 
 Richter (1792, v.1, p.xxix), 
 as quoted in Szabadv\'{a}ry (1966, p.102).  
 \end{quotation}

 {\meugr These equations express compositional rules and  
 mass conservation laws for reagents and products 
 that can be written as simple linear equations}, see 
 Leicester and Klickstein (1963, p.205-208), Muir (1907, p.269-276),  
  and Szabadv\'{a}ry (1966, p.97-113).

	  \begin{figure}[tb] 
  \iflatextortf
   {} 
  \else   
   \centerline{  
     \includegraphics[height=1.7in, width=1.1in,  
    trim=0.0in 0.0in 0.0in 0.0in, clip]{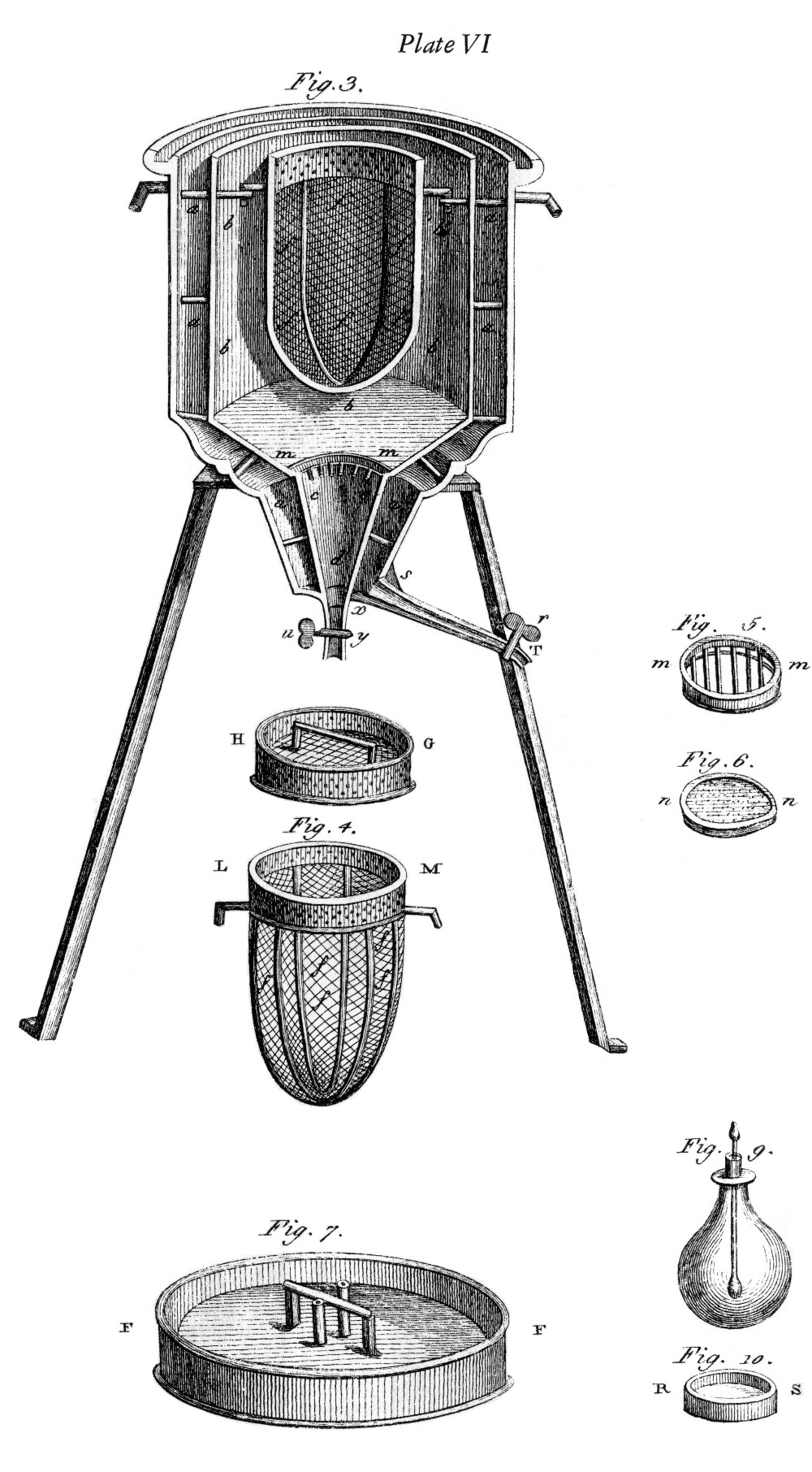} 
     \mbox{} \ \mbox{} 
     \includegraphics[height=1.7in, width=2.2in,  
    trim=0.0in 0.3in 0.0in 0.0in, clip]{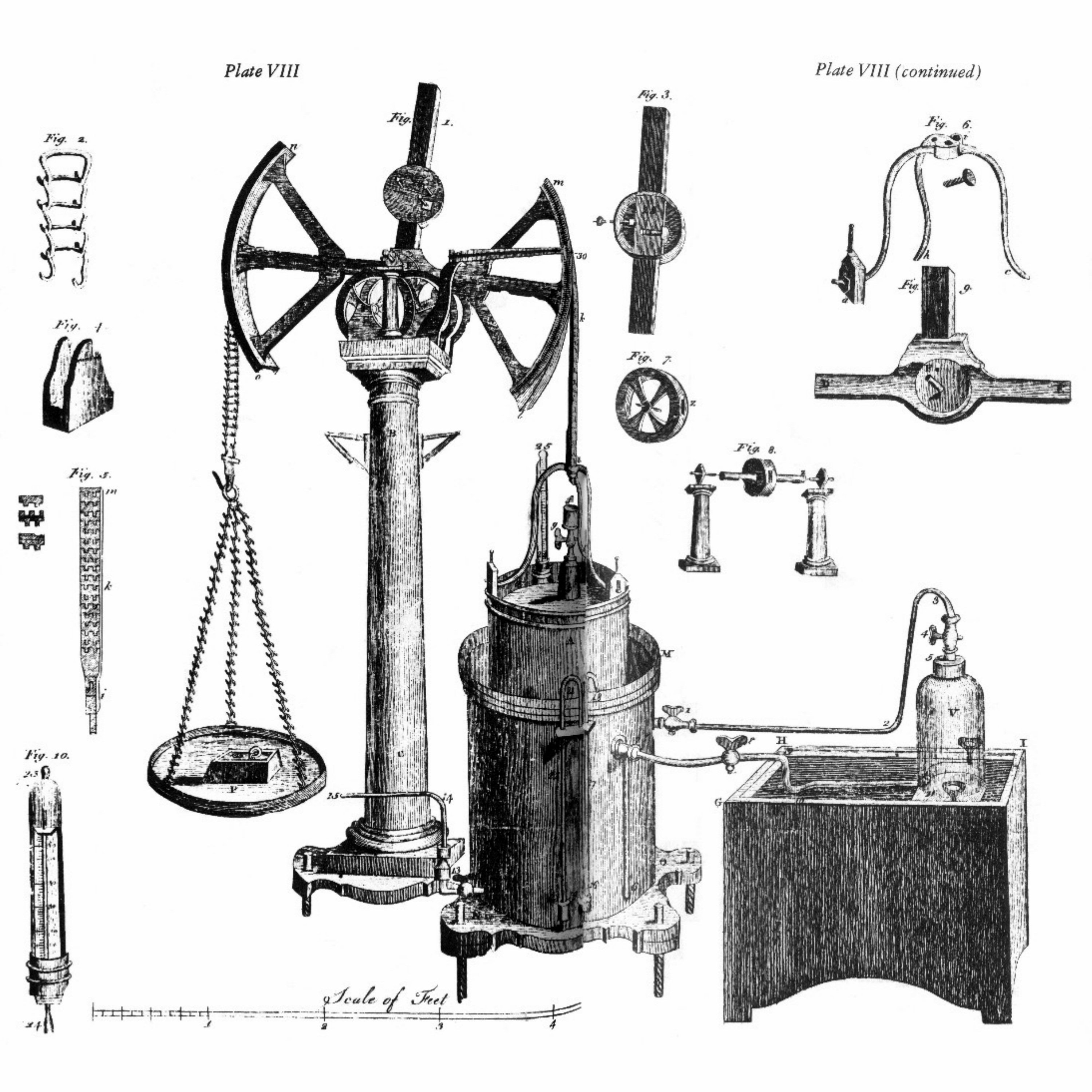} 
     \mbox{} \ \mbox{}  
     \includegraphics[height=1.7in, width=2.3in,  
    trim=0.0in 0.0in 0.0in 0.0in, clip]{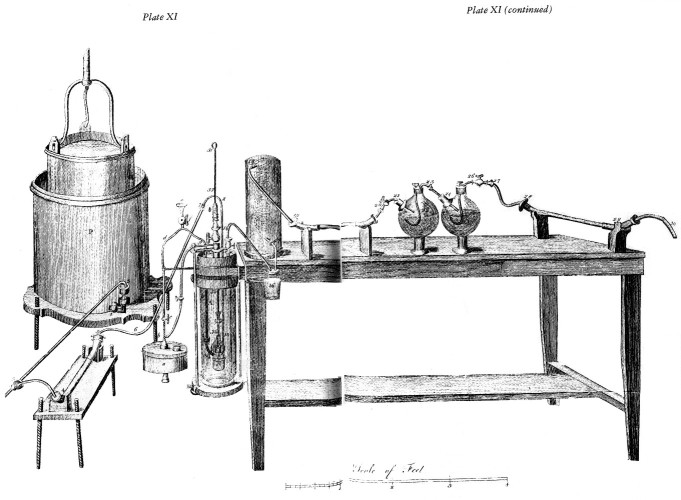} 
    } 
  \fi  
  \centerline{Figures 7abc: Lavoisier (1789, Plates VI, VIII and XI).} 
  \end{figure}

  The invariance properties and the conservation laws expressed by 
 stoichiometric balance equations mutually support each other. 
  Their validity can be jointly checked and confirmed by empirical 
 experiments based on careful mass measurements, made with the aid 
 of gravimetric (or, for gases, volumetric) instruments, 
 see Figure 7.     
  The relation between invariant elements of a theory and its 
 conservation laws is a very rich area of research, 
 that {\meuve I cannot} further explore in this paper.    
   {\meuve I will} return to this topic in following articles,      
 see Stern (2011b) for a more formal and detailed analysis.

   Chemistry can rely on stoichiometric rules, mass conservation laws,  
  and the associated invariant elements to achieve an orderly structure.  
   However, chemical theory will not be very useful if it does not 
  allow a practitioner to foresee, among all the stoichiometrically 
  well balanced equations, which ones correspond to chemical reactions 
  that actually occur.  
   These predictive powers concern the topic of chemical affinity, 
  to be addressed in the next section.

 \section{Ethics, Recycling and Affinity}  
   
  The ethical imperative, as discussed in Section 1, 
 requires innovation, that must however be conciliated with the 
 need of an autonomous system to preserve its kernel identity.  
  New and better opportunities must to be created, 
 conserving however the system's essential form and attributes, 
 preserving core mechanisms of its fundamental way of life. 
  Theoretical models and hypothesis, technological devices, 
 laboratory apparatus, data analysis algorithms and others means and 
 methods involved in the life of an autonomous scientific system 
 can all be replaced, as long as they are substituted by 
 compatible analogs, that is, by functional equivalents or 
 succedanea that allow the system to recycle its 
 {\it \meuve modus operandi}   (to maintain its autopoiesis).   
  Of course, accomplishing all that is no easy task. 
  Conflicting demands and inconsistent conditions must be dealt with, 
 difficult (ethical) choices need to be made.   
  This task always involves non-trivial, and sometimes profound,   
 systemic re-organizations or, using Piaget's nomenclature, 
 the re-equilibration of cognitive structures.
  In this section {\meuve I study} some ethical aspects of the 
	chemical revolution.

   The European corpus of chemical knowledge in the Middle Ages 
  is known as alchemy, inheriting traditions having their 
  roots in several ancient civilizations, that were collected, 
  reorganized and consolidated mostly by Arab sages. 
   Its language sounds strange to a modern scientist. 
   Alchemical texts explain how substances live and die  
  (or even, sometimes, resurrect).
   Alchemy explains the way chemical substances interact 
  telling stories about their passions and desires, how they 
  love and hate each other; wherefrom the origin of the term 
   {\it chemical affinity} can be traced. 
   Alchemical practice may call for intervention of the  
  supernatural or manipulation of hidden or occult forces. 
   Its texts often employ arcane symbology to convey its mysteries, 
  and are  intentionally written so to be obscure, because the 
  secrets they contain are not supposed to become widely known. 
   Finally, it seems that the magical-vitalistic nature of alchemical 
  explanation is essentially qualitative, having for a long time  
  precluded, or at least strongly discouraged, the development and 
  application of quantitative research methods, 
  see Alfonso-Goldfarb (2001, Ch.5), Crosland (1962, Ch.I-III), 
  Hansen (1978, p.484-490) and Sevcenko (2000).  
 
   At the beginning on the XVI century, the alchemical and related 
  literature had expanded to incorporate practical knowledge 
  of many areas, including: mining and metallurgy;  
  extraction and distillation of essences; preparation of 
  cosmetics and medicines; and production of a great variety of 
  materials, from glass to gunpowder. 
   Nevertheless, all these applications employed a basic set of 
  chemical substances, sharing common means and methods of 
  preparation.   
   The term {\it Stahlian chemistry} is often used referring to 
  pre-Lavoisier chemistry, due to the influence of 
  Georg Ernst Stahl (1659-1734).  
   However, notwithstanding all the practical progress attained 
  at this period, the old concepts of alchemical theory were 
  still very much alive, see Bensaude-Vincent (1991, p.419), 
  Leicester and Klickstein (1963, p.58-63), and 
  Lewowicz (2011, p.438).

   Inevitably, those working or otherwise interested in the field, 
  started to search for principles of rational explanation that 
  could help them to effectively organize and more easily 
  understand the subject of chemistry. 
   According to several researchers in history of chemistry, 
  the first satisfactory of such organizing principles 
  was chemical affinity, see  Chalmers (2012), 
  Goupil (1991, L.I-II), Klein (1996) and 
  Duncan (1996, p.27-29, 156-159, 223-225).

    In 1718, \'{E}tienne Fran\c{c}ois Geoffroy published his 
   {\it Table of Chemical Affinities}, 
   see Leicester and Klickstein (1963, p.67-75) 
   and Figure 9 in Appendix A.  
    Each column in this table is organized as a list, having 
   at the head an important chemical substance. 
    The list's tail contains substances able to combine with 
   the one at the head, ordered by decreasing chemical affinity. 
    Affinities, or tendencies to combine, are ranked according to  
   displacement reactions.  
    Geoffroy explicitly chooses the word {\it affinity}  (rapport) for 
   its neutral character, trying to stay away from vitalistic 
   interpretations and convey objective empirical information.

    The reception of Geoffroy work in 1718 was enthusiastic, 
   as one can tell from the first quotation that follows, 
   a contemporary report from Fontenelle, the historian of the 
   French Academy.    
    As one can see from the second quotation, three decades later  
    (1749),  M.Clausier saw Geoffroy work as the major organizing 
   principle of chemical science, and was ready to elevate affinity 
   relations to the status of {\it axioms of chemistry}.  
    Nevertheless, before the end of the century Lavoisier expressed 
   a much somber view, as one can see from the third and forth 
   quotations.  
    It seems that, in Lavoisier's melancholic judgment, 
   chemical affinity could become an important scientific concept 
   but, so far, it stood  on very shaky grounds.

  \begin{quotation} 
  {\myqt  It is here that the sympathies and attractions would 
	 come very much to the point, if they meant anything. 
   But in the end, leaving as unknown that which is unknown, and 
  keeping to certain facts, all the experiments of Chemistry prove 
  that a particular Substance has more disposition to unite with 
  one Substance than another, and that this disposition has 
  different degrees...  
   This Table becomes in some sort prophetic, because if substances 
  are mixed together, it can foretell the effect and result of 
  the mixture... 
    If Physics could not reach the certainty of Mathematics, 
   at least it cannot do better than imitate its order. 
    A Chemical Table is by itself a spectacle agreeable to the 
   Spirit, as would be a Table of Numbers ordered according to 
   certain relations or certain properties.} 
   Fontenelle (1718), as quoted in Duncan(1996, p.157).   
  \end{quotation}

  \begin{quotation} 
  {\myqt  The affinities as we give them here are only an assemblage 
   of experiments which having been repeated many times on the same 
   materials, are as good as axioms in Chemistry...}
   Clausier (1749, p.5), as quoted in Duncan(1996, p.157).  
  \end{quotation}

  \begin{quotation} 
  {\myqt  The part of Chemistry most susceptible, perhaps, of becoming 
  one day an exact science, is that which deals with affinities or 
  elective attractions. M.Geoffroy, M.Gellert, M.Bergman, M.Scheele, 
  M.de Morveau, M.Kirwan and many others have already assembled a 
  multitude of particular facts, which now await only the place which 
  has to be assigned to them; but the principal data are lacking, 
  or at least those which we have are not yet either precise enough 
  or certain enough to become the fundamental basis on which must 
  rest so important a part of Chemistry.}
  Lavoisier (1789, v.I, p.xiii-xiv), as quoted in Duncan(1996, p.158). 
  \end{quotation}

 \begin{quotation}
 {\myqt  This science of affinities, or elective attractions, holds the 
 same place with regard to the other branches of chemistry, as the 
 higher or transcendental geometry does with respect to the simpler 
 and elementary part; and I thought it improper to involve those 
 simple and plain elements, which I flatter myself the greatest 
 part of my readers will easily understand, in the obscurities and 
 difficulties which still attend that other very useful and 
 necessary branch of chemical science.  

  Perhaps a sentiment of self-love may, without my perceiving 
  it, have given additional force to these reflections.
  M.de Morveau is at present engaged in publishing the article
  {\em Affinity} in the Methodical Encyclopedia; and I had more 
  reasons than one to decline entering upon a work in which he is 
  employed.} 
  Lavoisier (1790, p.xiv,xxxvii). 
 \end{quotation}

   In light of the previous sections of this article, one can explain  
  the decline of prestige or decay in scientific status suffered by 
  the concept of chemical affinity as a consequence of the weak 
  aesthetical properties of its object of study, relative to the 
  new objects of stoichiometric analysis. 
   Geoffroy's affinity tables encode ranking relations that can 
  be translated as inequalities, not equations. 
   The article of Morveau cited by Lavoisier wraps affinity 
  relations in an elegant additive algebraic structure, 
  see Morveau et al. (1803, p.399-401),  
  Goupil (1991, p.179-189) and Appendix A.    
   Nevertheless, even using Morveau's formalism, experimental data 
  in the form of occurrence or non-occurrence of displacement 
  reactions can only render mathematical inequalities. 
   Hence, this setting cannot define precise values for chemical 
  affinities, but only enforce finite interval bounds.
    
   Morveau may have been way ahead of his time, see Figure 8.    
   Modern approaches to these issues describe equilibrium points in 
  networks of reversible reactions, not irreversible displacements, 
  see Stern and Nakano (2014).  
   In this context, Th\'{e}ophile Ernest de Donder (1936) 
  could derive, in 1923, affinity related eigen-solutions 
  exhibiting the exact additive structure foreseen by Moereau, 
  see Goupil (1991, L.III, Ch.4, Sec.4, p.297-305).   
   However, the notions of reaction velocity and state equilibrium 
  were only introduced by the work of Guldberg and Waage (1879)
  in the late 19th century, see Leicester and Klickstein 
  (1963, p.468-471) and Muir (1907, Ch.XIV, p.402-405).  
   It is interesting to note that Guldberg and Waage had the sense  
  of being on a mission to rescue the notion of chemical affinity 
  from its steady decline, stating that  
  \begin{quotation} 
 {\myqt  Although we have not solved the problem of chemical affinities,
   we think that we have indicated a general theory of chemical 
   reactions, namely, those wherein a state of equilibrium is 
   produced between opposing  forces... 

  All our wishes would be accomplished if we could succeed in drawing
 the serious attention of chemists, by means of this work, to a branch
 of chemistry which has undoubtedly been much neglected since the
 beginning  of the century.}

  Guldberg and Waage (1879), as quoted in Muir (1907, p.405). 
   \end{quotation} 

   Once we understand the reasons for the decline of affinity 
  at the time of Lavoisier, an ethical question arises: 
   What should have been the destiny given to all the knowledge 
  accumulated by Sthalian chemistry, a corpus that had been 
  organized around the decaying concept of chemical affinity? 
   Should all that knowledge have been scraped? 
   According to the notion of ethics formulated in Section 1 
  and in the first paragraph of the current section, 
  such a wasteful attitude should be considered profoundly 
  unethical! 

  In 1787, an elite group of French chemists, including 
 Guyton de Morveau, Antoine-Laurent de Lavoisier, 
 Claude-Louis Berthollet and Antoine Fourcroy, 
 introduced a new method of chemical nomenclature;  
 see Lavoisier et al. (1787) and 
 Leicester and Klickstein (1963, p.180-192).      
   
  Many researchers have investigated how the new language they 
 carefully crafted faithfully expresses and reliably encodes   
 the analytical mechanisms of Lavoisier's new chemistry, 
 see for example Beretta (1993, p.200), Crosland (1962, p.177-180),  
 Donovan (1993, p.70-71,308), Levere (2001, p.69-71), 
 and Poirier (1993, p.197). 
  {\meuve Perhaps willing to transmit the accuracy of the 
	new system, the aforementioned authors of the new method 
	of chemical nomenclature simply stated, 
	 as one of its basic principles, that:}      
 \begin{quotation} 
  {\myqt  Denominations should, as far as possible, 
   conform to the nature of things.}   \\ 
	 {\meuve Leicester and Klickstein (1963, p.182).}  
 \end{quotation}  

  Such a blunt statement may bring the danger of 
 {\it reification}, the na\"{\i}ve tendency, typical of 
 strong-realistic epistemological frameworks, to assume that a 
 given object in a cognitive domain concretely exists out-there  
 as a {\it Ding an sich} (a thing in itself).  
  From the way {\it objects} are constructed and {\it objectivity} 
 is characterized, the epistemological framework of cognitive 
 constructivism automatically shields the scientist from this kind of 
 misunderstanding;  for a more detailed discussion of this issue, 
 see Stern (2007b, 2011a,b).  
   
  In contrast to examples possibly perceived as reification 
 statements, the aforementioned statements of 
 Lavoisier (1790, p.xiii-xxxvii) exemplify a process of de-reification. 
  Affinity, the central concept of Stahlian chemistry, is qualified 
 with the adjectives (im){\it precise}, (un){\it certain}, 
 {\it obscure} and {\it transcendental}. 
  Such reification and de-reification movements are typical of 
 paradigm shifts in the historical development of scientific 
 disciplines.  
  
  In light of our discussion, it should be clear how the perceived 
 strength or weakness of general aesthetical properties of objects 
 of knowledge, that is, {\meuve the relative precision of invariance 
 relations and the power of compositional rules of the eigen-solutions} 
 they stand for, are eventually translated into reification or 
 de-reification statements.

  Many fewer researchers have investigated the success of 
 Lavoisier's new language in efficiently recycling an important 
 inventory of chemical substances and preparation methods used 
 in Stahlian chemistry, see for example 
 Donovan (1993, Ch.4) and Holmes (1989, p.i-ii,55,122). 
  As noted by J.L. Heilbron,      
 \begin{quotation} 
  {\myqt 
  [Holmes'] prime example is the chemistry of salts formed from 
	mineral acids and various bases. He shows that Lavoisier 
	{\tt took over this body of knowledge virtually intact}:     
  it had {\tt its own principles and logic}; and did not require 
 dephlogistication to be useful.}  \      
 Holmes, (1989 p.i-ii, emphasis mine).
 \end{quotation}  
  Eklund (1975) compiles a dictionary that re-presents old objects 
 using the new nomenclature. 
   {\meuve I should} stress that, in such a dictionary, 
 translation assumes some form of functional equivalence or 
 compatibility, it does not imply identity. 
   {\meuve I will} return to these issues in Section 6 and following 
	articles, studying the possibility and meaning of diachronic 
	ontological alignments.

 \section{Metaphysical Blind-Spots} 
         
  \begin{quotation} 
  {\myqt 
	
	 \centerline{Spectroscopy and the Chemists: A Neglected Opportunity?} 

   `It is remarkable for how long chemists neglected the precious
 mean of discrimination at their hands in the use of the prism
 --- a striking example of how much may be lost by a too exclusive
 devotion to one branch of science to the neglect of others.'    

  This observation, made by George Stokes (1885, p.34-35),
 echoes a number of earlier remarks on the delayed acceptance of
 spectrum analysis by chemists.

  An opinion similar to that held by Stokes was expressed by
 Herschel in a letter to John Tyndall, in which he stated that
 the work he had done in the 1820's
   `might have led an enquiring
 person at an earlier period than the present to much that in now
 strikingly brought out.'}  \ Sutton (1976, p.16). 
  \end{quotation}

   This is how M.A. Sutton  starts his article about
  spectroscopy as a long  {\it neglected opportunity} in
  chemical analysis. 
   The starting quotation by George Stokes highlights the lost 
  opportunity of (once again) interconnecting two well established 
  systems of human knowledge, namely, chemistry and physics. 
   As for the usefulness or even the need of this new technology, 
  Sutton (1978, p.19) quotes William Huggins (1899, p.5-7), 
  who compared it to
  ``the coming upon a spring of water in a dry and thirsty land''.  
   Sutton (1976, p.17) speculates on the factors that could have
  prevented an earlier adoption of spectroscopic methods in
  chemical analysis, suggesting, among the possible reasons, the 
   ``unavailability of the necessary apparatus'' and 
    ``suspicion of the consistency of the effect in the 
    absence of any adequate theory of its cause'', 
    see also Pearson and Ihde (1951) and Koirtyohann (1980).
   
  A brief examination of the selected landmarks in the succinct 
 time-line of spectroscopy in Section 2, summarized in Figure 8, 
 should be enough to dismiss the unavailability of good equipment as 
 a possible reason. 
  In fact, it is possible to completely invert this argument! 
  According to McGucken (1969, p.9,28-29), the greatest obstacle 
 hindering the development of spectral analysis was 
 the complexity of observed spectra. 
  Fortunately for Kirchhoff and Bunsen,  
 the limitations of the equipment they were using compelled them 
 to work with over-simplified data, containing  
 {\meuve ``only the more conspicuous characteristics of a spectrum''}. 
  On the one hand, such an over-simplified description was still 
 sufficient to distinguish chemical elements and, on the other 
 hand, allowed them to escape the complexity conundrum.

   The second reason stated by Sutton as an impediment for 
  the development of spectroscopic analysis, namely, 
  {\meuve ``suspicion of the consistency of the effect in the 
   absence of any adequate theory of its cause''},  
  must also be false, because the first acceptable physical model 
  for the line spectra of hydrogen, the simples of all atoms, 
  was only given in 1913 by Niels Henrik David Bohr (Figure 3a),   
  see  Bohr (1913, 1987, 1998), Einstein (1905),   
  Enge et al. (1972, Sec.4.7, p.99-104),   
  Stern(2008b, Ch5) and    
  Tomonaga (1962, v.I, Sec.3.18, p.97-106).  
   However, {\meuve I believe} it is possible to uphold a relaxed version 
  of this statement. 
   Before giving my own version, let us hear the opinion of 
   Gustav Robert  Kirchhoff, who is universally acknowledged 
  as the sole responsible for the final breakthrough that 
  triggered the spectroscopy revolution. 
  \begin{quotation} 
 {\myqt  I also have a few points to mention concerning the history 
of the chemical analysis of the solar atmosphere.
 The core of the theory of solar chemistry that I have developed 
consists of a proposition that, shortly stated, says: 
  For each kind of (heath or light) rays, the relation between the
 emission power and absorption power is the same 
 {\em (das gleiche)} for all bodies at the same temperature.    
  From this proposition it easily follows that a glowing body 
 that only emits light rays of certain wavelengths, 
 also only absorbs light rays of the same wavelengths;  
 wherefore it is revealed how it is possible to know 
 the constituents of the solar atmosphere 
 from the dark lines of the solar spectrum.} 
 Kirchhoff (1863, p.102-103).  
	\end{quotation}

  What makes this historical example so interesting 
 is that Kirchhoff did not provide an adequate theory for 
 the cause of spectral emission and absorption. 
  As stated in Section 1, Balmer's formula was the first available 
 explanation for the relative positions of (some) lines in atomic 
 spectra. 
  Kirchhoff did not even provide such an empirical recipe, 
 explaining nothing of the structural order found in spectra. 
  Instead, it seems that Kirchhoff provided the bare minimum for  
 the breakthrough, namely, a firm handle to grasp new objects, 
 an anchor securing a fixed point of view and allowing one 
 to stare straight into the phenomena at hand, 
 see Kirchhoff (1860b).    
  This anchor had the form of a simple law of equilibrium, having 
 nothing to do with specific models for atomic or spectral structure.  
  Instead, it took the form of a general thermodynamic equality 
 ({\it Gleichheit}) constraint. 
  {\meuve This opinion is supported by}    
 James (1983, p.42),    
 Rosenberger (1890,v.3, p.691-692) and 
 Schirmmacher (2003, p.299-301). 
 
  The minimality of Kirchhoff's explanation and the intensity of the 
 following revolution highlights a  {\it seeding effect} that seems 
 to be characteristic of the metaphysical imperative.  
  The presence of tiny particles in a dense cloud or in a 
 saturated solution have the power of triggering phase-transitions  
 in the form of wide spread condensation, precipitation or 
 crystallization.  
  In the same way, Kirchhoff's minimalist explanation sparked an 
 abrupt and radical transformation in its cognitive field. 
  Nevertheless, before the seeding, nothing happens, 
 the world stands still, see Jastrow (1899).  
  Heinz von Foerster (1995) used to convey this idea with his  
	\begin{quotation} 
	\noindent 
  { \it Principle of the Double Blind: The blind spot: 
        One does not see what one does not \nolinebreak see.}  
  \end{quotation}

 \section{Therapy: Curing the Blind}

   The therapeutic imperative seeks cure for systemic blindness,      
  revealing what is concealed, displaying what is occult.  
   However, as noted in Section 1, in order to introduce new objects 
 in a (upgraded) scientific ontology, one has to reach a higher level 
 in the scientific discipline (under construction), revisiting the 
 aesthetical, ethical and metaphysical imperatives. 
   The therapeutic imperative points to evolution, 
  seeking sharper images of an expanded reality, 
  looking for better scientific understanding of new phenomena, 
  by (typically, but not necessarily in that exact order) 
  speculating on alternative approaches and 
  investigating innovative hypotheses,  
  developing conventional means and methods and     
  coherently incorporating unconventional ones, 
  searching for new or sharper (eigen) solutions and, if necessary, 
  integrating all of the above in better theoretical frameworks.  
   This section shows how authentic therapeutic change is possible, 
  using two historical examples to illustrate this process.

 \subsection{From Vapor-Ware to Material States}

  The {\meuve first historical example in this section} is centered on 
 Lavoisier's new theory  of the gaseous state, the equipment he 
 developed to check mass conservation in chemical reactions, 
 see Figure 7,  
 and the emergence of the modern notion of chemical element.  
  The path followed by Lavoisier in the critical years of this 
 revolution constitute a complex story\footnote{ 
  For a lively account of this story, see Donovan (1993, Ch.4, p.74-109); 
  for interesting details on the evolution of measurement techniques and 
 devices, see Holmes and Levere (2002).}. 
  Many research topics and treads of development are intertwined 
 in this path including, among others: 
   Discussion of several conceptions of the classical five, four 
  (or even two) alchemical elements (a.k.a substantial principles 
  or natural instruments); 
   study of boiling or evaporation of liquids, 
  in open air or in evacuated chamber and  
  expansibility properties of several substances;   
   study of fixation and release of water in formation of salts 
  and crystallization phenomena; 
   study of fixation and release of airs in combustion and 
  calcination phenomena; 
   development of special balances, densimeters, pneumatic 
  equipment and other measurement devices;  
   development of a new theory of chemical combustion reactions,  
  new notion of physical states of matter, and 
  new definition of chemical elements. 

   In spite of all the inherent complexities of any real historical 
  process and the many works of Lavoisier, 
  Gough (1988, p.31) gives the following simplified outline 
  of Lavoisier's core steps for the   
  (Gough disapproves of the term {\it revolution}) momentous development 
  of chemistry:        
   \begin{quotation}  
   {\myqt  I should like to propose that Lavoisier brought into [a] 
 basically Stahlian framework three important and interrelated ideas.
 \begin{itemize}\iflatextortf{}\else{\itemsep2pt \parskip0pt \parsep0pt}\fi    
 \item 
 First of all (both historically and logically), Lavoisier formulated
a theory of the gaseous state of matter that allowed him to conceive of
numerous chemically distinct substances in an {\it a\"{e}riform} state as gases
rather than {\it airs}.  
 \item 
  Second, the realization that invisible aeriform fluids could leave
and enter substances during the course of chemical reactions
necessitated a gravimetric accounting to determine chemical composition.
 It was the systematic application of an unarticulated gravimetric
criterion of composition that allowed Lavoisier to determine the proper
order of chemical simplicity.
 \item 
  Finally, the fact that many a\"{e}riform substances existed in physical
states that were nearly identical, but in chemical states that were
quite divergent, prompted Lavoisier to apply the Stahlian reactive
criteria of chemical identity in a more thorough, rigid, and 
systematic fashion than ever before. 
 \end{itemize}     }
   \end{quotation}

   These three steps in Gough's outline are a perfect match to  
  our metaphysical, aesthetical and ethical concerns. 
   Speculating on several ideas about the nature of a\"{e}riform 
  {\meuve (air-like)} substances, Lavoisier could acknowledge the new 
	variables that he needed to incorporate in his measurements even if, 
  at the beginning, he could only faintly notice or distinguish 
  the corresponding entities.   
   Developing and building the necessary equipment, including some 
  adapted devices borrowed from the physics laboratory, 
  he could take accurate measurements, showing the nice aesthetical 
  properties (precision and compositional rules) of the newly 
  found invariant objects and conservation laws.    
   Using (informally) the (implicit) relation between conservation 
  laws and invariants of a theory, he used these invariant objects 
  to give a new characteriztion of chemical elements.  

   Finally, if Lavoisier was a revolutionary, 
  his revolution was a very ethical one. 
   Lavoisier {\meuve (1772)} may even have said, 
	 as quoted in Donovan (1993, p.104),  
   \begin{quotation} 
    {\myqt I have felt no obligation to consider anything 
     done before me as more than a hint.}      
   \end{quotation} 
   Nevertheless, from the metaphysical imperative,  
  we know very well how valuable a good hint can be.   
   Furthermore, the newly developed chemical theory was able 
  to efficiently recycle the important inventory of Stahlian chemistry, 
  doing so in a way that was consistent and coherent with the 
  basic organizing principles of the old framework.   
    {\meuve I will} further explore possible interpretations of the last 
  statement in following articles, as indicated at Section 1.

  \subsection{The Strange Case of Balmer's Formula} 
   
  As {\meuve a second historical example in this section} 
 let {\meuve us} examine Balmer’s formula,  a case that exhibits 
 some aspects often perceived as strange, exotic, or at least as 
 contrasting with the first historical example. 
  {\meuve We use} the following descriptive labels for future reference.
  \begin{description}\iflatextortf{}\else{\itemsep2pt \parskip0pt \parsep0pt}\fi   
   \item[Exogenous] aspect:  
    Balmer's intuition, allowing him to see the patterns he then 
  described using his celebrated empirical formula, 
  was based on models written in the language of projective and 
  descriptive geometry and embedded in the context of classical 
  architecture, see Banet (1966, p.503). 
   Such language and context are considered disconnected from   
  chemistry, hence perceived as exogenous, extraneous or exotic. 
   \item[Asynchronous] aspect:  
    The models used by Balmer, based on ancient geometry, 
   and the target application in spectroscopy, 
   seem to be unrelated in time, that is, there is no apparent 
   connection in the event chronology of the two fields.            
   \item[Disengagement] aspect:   
   Lavoisier's work instantly triggered a revolution in chemistry,  
  as did Kirchhoff's work in spectroscopy, see Section 2.  
   Meanwhile, the most important consequences of Balmer's work 
  had to wait for 30 years when, reinterpreted through the work of 
  Niels Bohr, it had a large impact in the quantum mechanics revolution.      
  \end{description} 
      
    {\meuve I am} intrigued by these strange aspects,  
  and tempted to ask the following questions:  
  \begin{itemize}\iflatextortf{}\else{\itemsep2pt \parskip0pt \parsep0pt}\fi 
   \item 
    Did {\meuve I choose}  {\it fair} descriptive labels for of these aspects? 
   \item 
    What are their conceivable causes, interconnections and  
    possible consequences?  
   \item 
    Could Balmer's formula have been discovered 
   before the statement of Kirchhoff's law? 
   \item 
    If so, would Balmer's formula be able to replace Kirchhoff's law 
   as a trigger for the spectroscopy revolution? 
  \end{itemize} 

   Any definitive answer to these question is in danger of being a 
  {\it nunc pro tunc} (now for then) conclusion, composing an 
  ill-conceived, goal-directed, trans-historical or 
  whiggish narrative. 
   Nevertheless, {\meuve I still} dare to propose these questions 
	for discussion in the spirit of 
  a Talmudic {\it pilpul}, that is, 
  a dialectical disputation aiming to examine the strength and 
  weaknesses of possible arguments or their constituent parts.

 \section{Future Research and Final Remarks}


  The present article explores the construction and evolution 
 of scientific ontologies. 
  One of the main objectives of future articles is to use the 
 epistemological framework of cognitive constructivism to further study 
 ontology alignments, focusing the question: How is it  possible to 
 coherently relate concepts and communicate  between different 
  worlds?   
  This is a recurrent problem in history and philosophy of science 
 referring to the possibility or impossibility of identifying   
 (or, in some reasonable sense, just matching) 
 entities belonging to two distinct theoretical frameworks.   
  Diachronic alignments concern specifically the case of 
 two theories developed at different epochs that, 
 nevertheless, aim to understand the ``same''  things.     
  This issue is closely connected to other 
 classical philosophical problems, including,    
  \begin{itemize}\iflatextortf{}\else{\itemsep0pt \parskip0pt \parsep0pt}\fi   
  \item  The problem of (in)commensurability of scientific 
   theories;      
  \item The problem of {\it objective symbol grounding} 
        in scientific ontologies. 
  \end{itemize}  
     
  The consecrated expression {\it external symbol grounding}    
 can still be used as long as we understand it as a reference 
 strictly external to language, because a symbol points to an 
 objective eigen-solution, but not as a reference to a 
 completely independent  {\it Ding an sich}\footnote{ 
  For influential philosophical discussions, see Feyerabend (1999) and 
 Kuhn (2000); for computer science and systems theory approaches 
 that will be instrumental in pursuing our goals, see  
 Feng et al. (2004), Goldstone and Rogosky (2002), and Harnad (1990).}.

  A second objective of following articles is to use the 
  epistemological framework of cognitive constructivism to study the 
 organization of science as a higher-order system having autonomous 
 disciplines as sub-systems.   
  This form of organization assumes the necessary and non-conflicting 
 features of operational closure and cognitive openness.  
  Such features can be enabled by coherent and harmonious communication 
 among distinct autonomous sub-systems based on  
 synchronic ontology alignments.     
  These issues concern the last of von Foerster aphorisms in Section 1, 
	the {\it organic imperative.}


  {\bf Acknowledgments:} 
  The author is grateful for the support of IME-USP  (The  Institute
 of Mathematics and Statistics of the University of S\~{a}o Paulo),
 FAPESP  (The State of S\~{a}o Paulo Research Foundation; 
 grant CEPID 2013/07375-0), 
 and CNPq  (The Brazilian National Counsel of Technological and 
 Scientific Development;  grant PQ 301206/2011-2).  
  The author is thankful for having had the opportunity to access the
 personal library of Prof. Haim Jurist,  
  the special library of History of Chemistry at PUC-SP 
   (The Pontifical Catholic University of S\~{a}o Paulo), 
 and the rare books and special collections section of the library of 
 IQ-USP (The Institute of Chemistry of the 
 University of S\~{a}o Paulo).
  The author is also grateful for the organizers and the constructive 
 criticism received from the following discussion fora: 
 UniLog-2013 (The 4th World Congress and School on Universal Logic); 
 ISPC-SS2013  (The Summer Symposium of the International Society 
 for the Philosophy of Chemistry); and   SIF-2013  
 (The XVII Congress of the Inter-American Society of Philosophy).  
  Finally, the author is grateful to Rafael Bassi Stern for his 
 critique of the first draft of this paper, to Jay Kadane for some 
 quick lessons on the art of bird watching at Itatiaia National Park, 
 and for valuable advice received from anonymous referees 
 and from CHK final editor, Jeanette Bopry, 
 whose suggestions were used to improve this work.



 \begin{figure}[tb] 
  \iflatextortf
   {} 
  \else   
   \mbox{} \  
   \begin{chronology}[2]{169}{192}{20pt}{16cm}
   \event{169.6}{Stahl}        
   \event{171.8}{Geoffroy}
   \event{174.9}{Clausier}     
   \event{177.3}{Lavoisier}    
   \event{178.5}{Morveau}      
   \event{179.3}{Richter}      
   \event{187.9}{Guldberg}  
   \event{192.3}{Donder} 
   \end{chronology}   
   \mbox{} \\ 
   \mbox{} \   
   \begin{chronology}[2]{169}{192}{20pt}{15.1cm} 
   \event{170.4}{Newton}
   \event{175.1}{Cronstedt}    
   \event{177.3}{Scheele}    
   \event{178.5}{Rittenhouse}  
   \event{180.2}{Wollaston}
   \event{181.4}{Fraunhofer}   
   \event{185.9}{Kirchhoff}    
   \event{188.5}{Balmer}       
   \event{190.5}{Einstein} 
   \event{191.3}{Bohr} 
   \end{chronology}   
   \mbox{} \\ 
  \fi 

 \vspace{2mm} 
 \centerline{Figure 8: Scientists' timelines  
             (approximate, by first work cited or published).}  
 \end{figure}  

\mbox{} 

\centerline{-------------------------------------------} 

\mbox{} 

\centerline{Some nice figures to fill this gap -- The M\"{o}bius band in Art and Technology} 
\centerline{Including:  R.Davis' noninductive and nonreactive resistor; G.Anderson's } 
\centerline{universal recycling symbol; a pentagonal paper folding;  and the  all-seeing}  
\centerline{Ouroboros in the Greek manuscript Chrysopoeia of Cleopatra, circa 100 C.E.} 

\mbox{}  

\centerline{ 
  \includegraphics*[height=1.2in,width=1.2in]{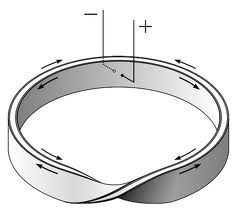} 
  \mbox{}  \  \  \  \mbox{} 
  \includegraphics*[height=1.2in,width=1.2in]{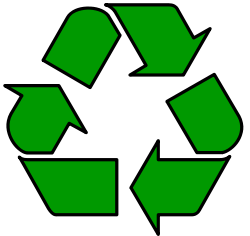} 
	} 

 \mbox{} 
	
	\centerline{ 
	\includegraphics*[height=1.2in,width=1.2in]{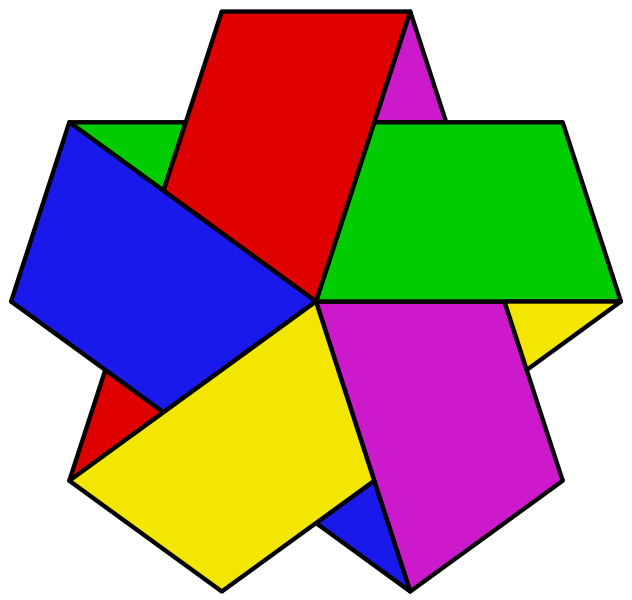} 
	\mbox{}  \  \  \  \mbox{}  
	\includegraphics*[height=1.2in,width=1.2in]{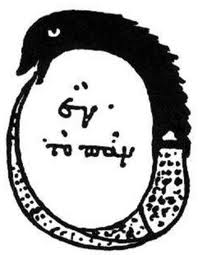}
	}

       
 \pagebreak

 \renewcommand{\baselinestretch}{0.90}
 \parskip 0.85mm

 \begin{footnotesize} 

 \section*{References} 

   \noindent 
  \rr Alfonso-Goldfarb,A.M. (2001). {\it Da Alquimia \`{a} Quimica}. 
  S\~{a}o Paulo: Landy. 

 \rr Balmer,J.J. (1885).
 Notiz \"{u}ber die Spectrallinien des Wasserstoffs. 
 {\it Annalen der Physik,} 261, 5, 80-87. 
 Translated as: Note on the Spectral Lines of Hydrogen. 
 in H.A.Boorse, L.Motz (1966). 
 The World of the Atom, V.1. New York: Basic Books. 
   
 \rr  Balmer,J.J. (1897). 
 Eine neue Formel f\"{u}r Spectralwellen. 
 {\it Annalen der Physik,} 296,2,380-391. 
 Translated as:  
 A New Formula for the Wave-Lengths of Spectral Lines. 
 {\it Astrophysical Journal}, 5,199-209. 

 \rr Banet,L. (1966). Evolution of the Balmer series. 
  {\it American J.of Physics,} 34, 496-503.

 \rr Banet,L. (1970). Balmers manuscripts and construction of his series. 
  {\it American J.of Physics,} 38, 821-828.

 \rr Bensaude-Vincent,B. (1991). Lavoisier: Una Revoluci\'{o}n 
  Cient\'{\i}fica. In M.Serres (Ed.). 
  {\it Historia de las Ciencias}. Madrid: C\'{a}tedra.

 \rr Beretta,M. (1993). {\it The Enlightenment of Matter: 
 The Definition of Chemistry from Agricola to Lavoisier.} 
 Uppsala Studies in History of Science, 15. 
 Science History Publications, USA. 


\rr Blackwell,D.; Girshick,M.A. (1954). {\it Theory of Games and 
   Statistical Decisions}. New York: Wiley.  
	 Reprint (1976) New York: Dover.  
	
  \rr Bohr,N.H.D. (1913). On the Constitution of Atoms and Molecules. 
  {\it Philosophical Magazine}, 
  Part I: 26, 1-24; Part II: 26, 476-501. 

 \rr Bohr,N.H.D. (1987, 1998).
    {\it The Philosophical Writings of Niels Bohr.}
    V.I (1987)- Atomic Theory and the Description of Nature.
    V.II and III (1987)- Essays on Atomic Physics and Human Knowledge. 
    V.IV (1998)- Causality and Complementarity.     
    Woodbridge, CN: Ox Bow Press.  

 \rr Border,K.C. (1989). {\it Fixed Point Theorems with Applications 
   to Economics and Game Theory.}  Cambridge University Press. 

 \rr Borges,W.; Stern,J.M. (2007). The Rules of Logic Composition for the
  Bayesian Epistemic e-Values. {\it Logic Journal of the IGPL},
  15, 5-6,  401-420.  doi:10.1093/jigpal/jzm032 .

 \rr Brier,S. (2001). Cybersemiotics and Umweltlehre. {\it Semiotica},  
 Special issue on Jakob von Uexk\"{u}ll's Umweltsbiologie, 
 134 (1/4), 779-814.

 \rr Brier,S. (2005). The Construction of Information and Communication: 
 A Cyber-Semiotic Re-Entry into Heinz von Foerster's Metaphysical 
 Construction of Second Order Cybernetics. 
 {\it Semiotica,} 154, 1, 355--399.   

 \rr Cerezetti,F.V. (2013). {\it Arbitragem em Mercados Financeiros: 
  Uma Proposta Bayesiana de Verifica\c{c}\~{a}o}. 
  Ph.D. Thesis, Institute of Mathematics and Statistics of the 
  University of S\~{a}o Paulo.    

 \rr Chalmers,A. (2012). Klein on the origin of the concept of chemical 
  compound. {\it Found.Chemistry}, 14, 37-53.  

  
  \rr Crosland,M.P. (1962). {\it Historical Studies in the Language of 
  Chemistry.} Harvard Univ.Press. 

 \rr DeBrander,F. (2007). {\it Spinoza and the Stoics.} 
     New York:  Continuum. 

 \rr Donder,T.E.de  (1936) {\it Thermodynamic Theory of Affinity: 
    A Book of Principles}. {\meuve Brussels: Stanford University Press. }       
    
 \rr Donovan,A. (1988). Lavoisier and the Origins of Modern Chemistry. 
   {\it Osiris,} 4, 214-231. 

 \rr Donovan,A. (1993). {\it Antoine Lavoisier: Science, Administration 
  and Revolution}. {\meuve Oxford, UK: Blackwell. }   
  
 \rr Donovan,A. (1990). Lavoisier as Chemist and Experimental Physicist: 
  Reply to Perrin. {\it Isis,} 81,2,270-272. 

 \rr  Dubins,L.E.; Savage,L.J. (1965). 
 {\it How to Gamble If You Must. 
 Inequalities for Stochastic Processes.}  
 New York:  McGraw-Hill.   


 \rr Duncan,A.M. (1996). {\it Laws and Order in Eighteenth-Century
    Chemistry}. Oxford: Clarendon Press. 

 \rr Einstein,A. (1905). \"{U}ber einen die Erzeugung und Verwandlung
 des Lichtes betreffenden heuristischen Gesichtspunkt (On a Heuristic
 Viewpoint Concerning the Production and Transformation of Light).
 {\it Annalen der Physik}, 17, 6, 132-148.

 \rr Eklund,J. (1975). {\it The Incomplete Chemist: 
  Being an Essay on the Eighteenth-Century Chemist in his Laboratory, 
  with a Dictionary of Obsolete Chemical Terms of the Period.}  
  Smithsonian Studies in History and Technology, 33, 1-49. 
  Washington, DC: Smithsonian Institution Press.   

 \rr Enge,H.A.; Wehr,M.R.; Richards,J.A. (1972).
  {\it Introduction to Atomic Physics.}  New York: Addison-Wesley.

 \rr Feng,Y.; Goldstone,R.L.; Menkov,V. (2004). ABSURDIST II:
  A Graph Matching  Algorithm and its Application to Conceptual
  System Translation. {\meuve pp.640-645 at  Barr,V.; Markov,A. (2004). 
	17th FLAIRS Conference.   Palo Alto, CA: AAAI Press. }  
     
 \rr   {\meuve Feyerabend,P.K. (1999). {\it Against Method: Outline of an 
      Anarchistic Theory of Knowledge}. London: New Left Books. }  

 \rr Finetti,B.de  (1972). {\it Probability, Induction and Statistics.} 
 New York: Wiley. 

 \rr Finetti,B.de  (1974). {\it Theory of Probability,} 
 V1 and V2. London: Wiley.  


 \rr   Foerster,H.von (1995). Anthology of Principles Propositions, 
  Theorems, Roadsigns, Definitions, Postulates, Aphorisms, etc. 
  Downloaded on 01/01/2013 from  
	{\tt http://www.cybsoc.org/heinz.htm} 

  \rr  Foerster,H.von (2003). {\it Understanding Understanding:
  Essays on Cybernetics and Cognition.} New York: Springer.   

   \rr {\meuve Fontenelle,B.B.de (1718). 
   Sur les Rapports de Differentes Substances en Chimie.
   {\it Histoire de l'Acad\'{e}mie Royale des Sciences}, 
	 1718, p.35-37.}   

  \rr {\meuve Geoffroy,E.F. (1718). Table des Diff\'{e}rents Rapports 
     Observ\'{e}s en Chimie entre Diffrentes Substances.   
     {\it M\'{e}moires de l'Acad\'{e}mie Royale des Sciences}, 
      p.202-212. }  
  
 \rr  Goldstone,R.; Rogosky,B. (2002). Using Relations within Conceptual
   Systems to Translate across Conceptual Systems.
   {\it Cognition,} 84, 295-320.  

  \rr Golinski,J. (1995). The Nicety of Experiments:
 Precision of Measurement and Precision of Reasoning in Late
 Eighteenth-Century Chemistry. p.72-91 in M. Norton-Wise (1995).
 {\it The Values of Precision.}  Princeton Univ.Press.

  \rr Gough,J.B. (1988). Lavoisier and the Fulfillment of the Stahlian
  Revolution. {\it Osiris}, 4, 1, 15-33.

  \rr Goupil,M. (1991). {\it Du Flou au Clair? Histoire de 
   l'Affinit\'{e} Chimique de Cardan \`{a} Prigogine}. 
     Paris: CTHS. 

   \rr Guerlac,H. (1961). {\it Lavoisier - The Crucial Year: 
   The Background and Origin of his First Experiments on 
   Combustion in 1772}. Cornell Univ. Press. 
		
  \rr Guldberg,C.M.; Waage,P. (1867). {\it \'{E}tudes sur les 
  Affinit\'{e}s Chimiques}. Oslo: Brogger and Christie.  

  \rr Hansen,B. (1978). Science and Magic. Ch.15, p.480-506  
    in D.C.Lindberg, {\it Science in the Middle Ages}. 
    Univ.of Chicago Press. 

 \rr Harnad,S. (1990). The Symbol Grounding Problem. 
    {\it Physica D}, 42, 335-346.
 	
 \rr Holmes,F.L. (1989). 
 {\it Eighteenth Century Chemistry As an Investigative Enterprise:  
 Five Lectures Delivered at the International Summer School 
 in History of Science Bologna.}  
 Berkeley Papers in History of Science, 
 Vol. 12. Berkeley, CA: Office for History of Science and Technology. 

  \rr Holmes,F.L.; Levere,T.H. (2002). {\it Instruments and 
  Experimentation in the History of Chemistry}. 
  Cambridge: The MIT Press. 

 \rr Ingrao,B.; Israel,G. (1990). {\it The Invisible Hand. Economic 
  Equilibrium in the History of Science.} Cambridge, MA: The MIT Press. 
 
 \rr Inhasz, R., Stern, J. M. (2010). Emergent Semiotics in Genetic 
  Programming and the Self-Adaptive Semantic Crossover.
  {\it Studies in Computational Intelligence}, 314, 381-392.
     
 \rr Inhelder,B.; Garcia,R.; Von\`{e}che,J. (1976). 
  {Epist\'{e}mologie G\'{e}n\'{e}tique et  \'{E}quilibration. 
   Hommage \`{a} Jean Piaget.} 
   Neuch\^{a}tel, Switzerland: Delachaux et Niestl\'{e}. 

  \rr James,F.A.J.L. (1983). The Establishment of Spectro-Chemical 
  Analysis as a Practical Method of Qualitative Analysis, 1854-1861. 
  {\it Ambix,} 30, 1, 32-53. 

  \rr Jastrow,J. (1899). The Mind's Eye.
  {\it Popular Science Monthly}, 54, 299-312.
  Reprinted in J.Jastrow (1900). 
  {\it Fact and Fable in Psychology.}
  Boston: Houghton Mifflin.   

 \rr Jech,Th.J. (1973). {\it The Axiom of Choice}. Amsterdam: 
    North-Holland. 

  \rr Jensen,W.B. (1986). The Development of the Blowpipe Analysis. 
	p.123-149 in Stock,J.T; Orna, M.V. 
  {\it The History and Preservation of Scientific Instrumentation,} 
  Dordrecht: Reidell.  

  \rr Kihlstrom,J.F. (2006). Joseph Jastrow and His Duck - 
	  Or Is It a Rabbit? On-line document, University of California 
		at Berkeley. Retrieved September 10, 2014 from \\ 
		{\tt http://socrates.berkeley.edu/$\sim$kihlstrm/JastrowDuck.htm}

  \rr Kirchhoff,G.R. (1860a). \"{U}ber die Fraunhofer'schen Linien. 
   {\it Annalen der Physik,} 185, 1, 1860, 148-150. 
    Transl: On Fraunhofer's Lines, 
    {\it Philosophical Magazine,} 19 (1860), 193-197.  

 \rr  Kirchhoff,G.R. (1860b). 
   \"{U}ber das Verh\"{a}ltnis zwischen dem Emissionsverm\"{o}gen 
   und dem Absorptionsverm\"{o}gen der Korper fur W\"{a}rme und Licht. 
   {\it Annalen der Physik,} 185, 2, 275-301. 
   Transl:  On the Relation Between the Radiating and 
   Absorbing Powers of Different Bodies for Light and Heat. 
   {\it Philosoph. Magazine,} 20, 1-21.
      
 \rr Kirchhoff,G.R.; Bunsen,R.W. (1860). 
    Chemische Analyse durch Spectralbeobachtungen. 
    {\it Annalen der Physik,} 186, 6, 161-189. 
    Translated as: Chemical Analysis by Spectrum-Observations. 
    {\it Philosophical Magazine,} 22 (1861), 329-249, 498-510. 

 \rr Kirchhoff,G.R. (1863). Zur Geschichte der Spectral-Analyse 
    und der Analyse der Sonnenatmosph\"{a}re. 
   {\it Annalen der Physik,} 194, 1, 94-111.
   Translated as: Contributions Towards the History of Spectrum 
   Analysis and the Analysis of the Solar Atmosphere. 
   {\it Philosophical Magazine,} 25, 250-262. 

 \rr Klein,U. (1996). The Chemical Workshop Tradition and the 
   Experimental Practice: Discontinuities within Continuities. 
    {\it Science in Context}, 9, 3, 251-287. 
  
 \rr Koirtyohann,S.R. (1980). 
    A History of Atomic Absorption Spectroscopy
    {\it Spectrochimica Acta,} 35B, 663-670. 

 \rr {\meuve Krohn,W., Kueppers,G., Nowotny,H. (1990). 
   {\it Selforganization. Portrait of a Scientific Revolution.} 
   Dordrecht: Kluwer.}    
  
  \r	Kuhn,T.S. (2000a). What are scientific revolutions? 
  p.13-32 in Kuhn et al. (2000). 

  \rr Kuhn,T.S. (2000b). Commensurability, Comparability, Communicability. 
  p.33-57 in Kuhn et al. (2000). 
	
	\rr Kuhn,T.S.; Conant,J.; Haugelan,J. (2000).  
   {\it The Road Since Structure}. University of Chicago Press.

  \rr  Lavoisier,A.L.de {\meuve (1772).} Memorandum of February 20. 
	     Reprinted in Guerlac (1961, p.228-230).  

  \rr {\meuve Lavoisier,A.L.de  (1774). 
  Opuscules physiques et chimiques (review). 
  {\it Histoire de l'Acad\'{e}mie Royale des Sciences,} 71-78. }  
	 Also in Lavoisier (1862). {\it Oeuvres}, v.2, p.89-96. 
	 Paris: Imprimerie Imp\'{e}riale.
   Lavoisier is identified as the author of this review in 
   {\it Oeuvres}, v.5 (1892), p.320. 

   \rr {\meuve Lavoisier,A.L.de  (1788). 
   M\'{e}moire sur la fermentation spiritueuse. 
   {\it Histoire de l'Acad\'{e}mie Royale des Sciences}.    
	   Reprinted in Lavoisier (1865). {\it Oeuvres.} 
     v.3, p.777-790. Paris: Imprimerie Imp\'{e}riale. }  

  \rr {\meuve Lavoisier,A.L.de  (1789). 
  {\it Trait\'{e} \'{e}lementaire de chimie, 
  p\'{e}rsent\'{e} dans un ordre nouveau et d'apr\`{e}s les
  d\'{e}couvertes modernes}. Paris: Chez Cuchet.    
	Translation: Kerr,R. (1790). {\it Elements of Chemistry}. }    
	  

 \rr  Leicester,H.M.; Klickstein,H.S. (1963). {\it A Source Book in
   Chemistry, 1400-1900}. Cambridge, MA: Harvard University Press.  

  \rr Levere,T.H. (2001). {\it Transforming Matter: 
	 A History of Chemistry from Alchemy to the Buckyball.} 
	 Baltimore, MD: The Johns Hopkins University Press. 

 \rr Lewowicz,L. (2011). 
   Phlogiston, Lavoisier and the Purloined Referent. 
   {\it Studies in History and Philosophy of Science,} 42, 436-444. 

 \rr {\meuve Luhmann,N. (1989). {\it Ecological Communication.} 
    Chicago Univ. Press.}    

 \rr  Madruga,M.; Esteves,L.; Wechsler,S. (2001).
 On the Bayesianity of Pereira-Stern Tests.
 {\it Test}, 10,291--299.
    
 \rr Mann,N.I.; Dingess,K.A.; Slater,P.J.B. (2006). 
  Antiphonal four-part synchronized chorusing in a Neotropical wren. 
  Biology Letters 2, 1-4. 

  \rr  Maturana,H.R.;  Varela,F.J. (1980).
 {\it Autopoiesis and Cognition. The Realization of the Living.}
 Dordrecht: Reidel.

 \rr McGucken,W. (1969). {\it  Ninteenth Century Spectroscopy. 
  Development of the Understanding of Spectra 1802 - 1897.} 
  Baltimore: Johns Hopkins. 

 \rr McMullin,B. (2004). Thirty Years of Computational Autopoiesis: 
   A Review. {\it Artificial Life}, 10, 3, 277-295.

  \rr McMullin,B.;  Varela,F.J. (1997). Rediscovering Computational 
	Autopoiesis. p. 38-47 in Husbands,P.; Harvey, I. 
  {\it 14th European Conference on Artificial Life}. 
	Cambridge, MA: The MIT Press.  

 \rr Melhado,E.M. (1985). Physics, and the Chemical Revolution.  
  {\it Isis,} 76, 2, 195-211.  

 \rr Meldrum,A.N. (1930). {\it The Eighteenth Century Revolution 
     in Science – The First Phase}. 
		 Calcutta: Longmans, Green and Co.

 \rr Morgenstern,O.;  Neumann,J.von (1947).  
 {\it The Theory of Games and Economic Behavior.}  
 Princeton University Press. 

 \rr Morveau, L.B.G.de; Maret, M.; Duhamel, M.  (1786).  
  {\it Encyclop\'{e}die M\'{e}thodique: Chymie, Pharmacie et 
   M\'{e}tallurgie}.  {\meuve Paris: Chez Panckoucke.}  
	 (Entry {\it Affinit\'{e}} at v.2, p.535-613) 
	Translated as: Affinity, p.391-405 in  
  {\it Supplement to the Encyclopaedia or Dictionary of Arts, Sciences 
   and Miscellaneous Literature.} Philadelphia: Thomas Dobson (1803). 
  
 \rr  Morveau, L.B.G.de;  Lavoisier,A.L.;  Berthollet,C.L.;  
   Fourcroy,A. (1787).  
 {\it M\'{e}thode de Nomenclature Chimique}.  
  {\meuve Paris: Chez Cuchet.}    
 Reprinted (1994) w.introduction by B.Bensaude-Vincent. 
 Paris: Ditons du Seuil. 
     
 \rr Muir,P. (1907). 
 {\it A History of Chemical Theories and Laws}. 
  New York: John Wiley.  

 \rr Partington,J.R. (1962). {\it A History of Chemistry}. 
  London: Mcmillan. 

 \rr Pearson,T.H.;  Ihde,A.J. (1951). Chemistry and the Spectrum Before 
  Bunsen and Kirchhoff. {\it Journal of Chemical Education}, 
  28, 5, 267-271. 


 \rr  Pereira,C.A.B.; Stern,J.M. (1999). Evidence and Credibility:
 Full Bayesian Significance Test for Precise Hypotheses.
 {\it Entropy Journal}, 1, 69--80.

 \rr Pereira,C.A.B.; Wechsler,S.; Stern,J.M. (2008).
 Can a Significance Test be Genuinely Bayesian?
 {\it Bayesian Analysis,} 3, 1, 79-100.

  \rr Perrin,C.E. (1984). 
  Did Lavoisier report to the Academy of Sciences on his own Book?
  {\it Isis,} {\meuve 75, 343-348.}   

 \rr Perrin,C.E. (1990). Chemistry as Peer of Physics: A Response to 
   Donovan and Melhado on Lavoisier. {\it Isis,} 81, 2, 259-270.  

 \rr Poirier,J.P. (1993). {\it Antoine Laurent Lavoisier: 1743-1794}. 
  Paris: Pygmalion
 

 \rr Rasch,W. (2000). {\it Niklas Luhmanns Modernity. Paradoxes of
 Differentiation}. Stanford Univ.Press. 

  \rr Richter,J.B. (1792). {\it Anfangsgr\"{u}nder der 
 St\"{o}chyometrie oder Messkunst chymischer Elemente.}   
 Breslau-Hirschberg: J.F.Korn.   

 \rr Rosenberger,F. (1890). {\it Geschichte der Physik  
   in Grundz\"{u}gen mit Synchronistichen Tabellen der Mathematik, 
  der Chemie ind Beschreibenden Naturwissenschaften sowie der 
  Allgemeinen Geschichte.}  
  Braunschweig: Friedrich Vieweg.  

 \rr Schirrmacher,A. (2003). Experimenting theory: The proofs 
   of Kirchhoffs radiation law before and after Planck. 
   {\it Historical Studies in the Physical and Biological Sciences}, 
   33, 2, 299-335. 

  \rr  Segal,L. (2001). {\it The Dream of Reality.
  Heintz von Foerster's Constructivism.}  New York: Springer. 

 \rr N.Sevcenko (2000). Apresenta\c{c}\~{a}o ou O Arco-\'{I}ris, 
    o Pote de Ouro e o Lago. p.11-14 in W.Weischedel (2000). 
    {\it A Escada dos Fundos da Filosofia}. S\~{a}o Paulo: Angra. 

 \rr  Shafer,G.; Vovk,V. (2001). {\it Probability and Finance, 
     It's Only a Game!} New York: Wiley. 
  
 \rr Shashkin,Yu.A. (1991). {\it Fixed Points}. 
    Providence, RI: American Mathematical Society.  
	
	\rr Spinoza,B. (1677). {\it Ethica: Ordine Geometrico Demonstrata.} 
   Reprint (2007) w. Portuguese translation, 
	 S\~{a}o Paulo: Aut\^{e}ntica.  
	
 \rr Stern,J.M. (2003).
  Significance Tests, Belief Calculi, and Burden of Proof in
  Legal and Scientific Discourse.  
  {\it Frontiers in Artificial Intelligence and its Applications,}
  101, 139--147. 

 \rr Stern,J.M. (2004). Paraconsistent Sensitivity Analysis for
 Bayesian Significance  Tests. 
 {\it Lecture Notes Artificial Intelligence,} 
 3171, 134--143.
 
 \rr Stern,J.M. (2007a).
 Cognitive Constructivism, Eigen-Solutions, and Sharp
 Statistical Hypotheses. {\it Cybernetics \& Human Knowing},
 14, 1, 9-36.

 \rr Stern,J.M. (2007b).  Language and the Self-Reference Paradox.
 {\it Cybernetics \& Human Knowing},
 14,4,71-92.

 \rr Stern,J.M. (2008a).  Decoupling, Sparsity, Randomization, and
 Objective Bayesian Inference.
 {\it Cybernetics \& Human Knowing}, 15,2,49-68.

 \rr Stern,J.M. (2008b). {\it Cognitive Constructivism and the Epistemic
 Significance of Sharp Statistical Hypotheses.} Tutorial book for
 MaxEnt 2008, The 28th International Workshop on Bayesian Inference
 and Maximum Entropy Methods in Science and Engineering.
 July 6-11 of 2008, Borac\'{e}ia, S\~{a}o Paulo, Brazil. 

 \rr Stern,J.M. (2011a). Constructive Verification, Empirical Induction,
 and Falibilist Deduction: A Threefold Contrast.
 {\it Information}, 2, 635-650

 \rr Stern,J.M. (2011b). Symmetry, Invariance and Ontology in Physics
 and Statistics. {\it Symmetry}, 3, 3, 611-635.

 \rr Stern,J.M.; Pereira,C.A.B. (2014). 
  Bayesian Epistemic Values: Focus on Surprise, Measure Probability! 
  {\meugr  {\it Logic Journal of the IGPL}, 22, 2, 236-254.}   

\rr {\meuve Stern, J.M.; Nakano, F.	 (2014). Optimization Models for Reaction Networks: 
    Information Divergence, Quadratic Programming and Kirchhoff's Laws. 
		{\it Axioms}, 3, 109-118.  {\tt doi:10.3390/axioms3010109}    } 

 \rr Stevens,A. (1999). {\it Ariadne's Clue: A Guide to the Symbols 
    of Humankind.} Princeton Univ.Press. 
   
 \rr Stevens,A.; Price,J. (2000). {\it Evolutionary Psychiatry: 
    A New Beginning}. London: Routledge. 

 \rr Stevens,A. (2003). {\it Archetype Revisited: An Updated Natural 
    History of the Self}. 
		Toronto: Inner City Books. 

  \rr Sutton,M.A. (1976). Spectroscopy and the Chemists: A Neglected 
  Opportunity? {\it Ambix,} 23, 16-26. 

 \rr Szabadv\'{a}ry,F. (1966). {\it History of Analytical Chemistry.} 
    London: Pergamon Press. 

 \rr Szab\'{o},A. (1978). {\it The Beginnings of Greek Mathematics}. 
    Budapest: Akad\'{e}miai Kiad\'{o}. 

 \rr Tomonaga,S. (1962). {\it Quantum Mechanics.  
  V.1 Old Quantum Theory; V.2, New quantum theory.} 
  {\meuve Amsterdam:  North Holland. }    

 \rr Varela,F. (1978). {\it Principles of Biological Autonomy.}
  North-Holland.

 \rr Wittgenstein, L. (1953). {\it Philosophical investigations}. 
    Oxford: Blackwell.

  \rr Zangwill,W.I.; Garcia,C.B. (1981).
   {\it Pathways to Solutions, Fixed Points, and Equilibria.}
   New York: Prentice-Hall.

 \end{footnotesize} 


 \pagebreak

 \setcounter{section}{0}
 \appendix

 \renewcommand{\baselinestretch}{1.0}
 \parskip 0.14cm

 \section{Morveau's Affinity Tables and Diagrams}

  For the reader's convenience, Appendix A presents some examples 
 of Morveau's reaction diagrams and a small affinity table, see 
 Morveau (1786, p.553,558,773), Morveau (1803, p.399-401),  
 and also Goupil (1991, L.II, Ch.V, Sec.4, p.179-189).  
  These examples should make clear how to compute, 
 from the affinity data, quiescent and divellent affinities.      
  If the former are smaller than the latter, the involved 
 displacement (irreversible) reactions are supposed to 
 occur\footnote{ \meuve  
  In contemporary chemistry, divellent and quiescent affinities, 
	also known as forward and reverse chemical affinities, are the 
	absolute values of stoichiometricaly weighted partial sums 
	 over products and reactants defining the vector of changes 
	 in chemical potential for a reaction network. 
	 The words divellent and quiescent are etymologically  derived from the 
	Latin verbs {\it divellere} - to separate, to tear apart - and {\it quiescere} 
	 - to remain quiet, to stay still. This etymological derivation conveys the 
	idea that forward reactions separate the constituent elements of 
	reacting compounds in order to form new products, while reverse 
	 reactions regenerate original reactants.  For further details, 
	 see Stern and Nakano (2014).    
  }.  
  In order to facilitate the reading of these diagrams, we also 
 provide  translations of chemical names and formulas, 
 even though fully aware that successive translations chains like:  
 \iflatextortf{  
  Vitriol of potash, Sulphat of potash, Potassium sulfate, $K2SO4$; 
  or 
  Mephite of barytes, Carbonate of barytes, Barium carbonate, $BaCO3$;  
  require non-trivial and consecutive diachronic ontological alignments. 
  The ionic valencies (of today's chemistry) involved in the next 
  reactions are as follows;   
  Cations:   
   Hydrogen, $H, \, {1+}$; Potassium, $K, \, {1+}$; 
   Ammonium, $(NH4), \, {1+}$; Barium, $Ba, \, {2+}$; 
   Calcium, $Ca, \, {2+}$; 
  Anions:  
  Hydroxide, $(OH), \, {1-}$; Chloride, $Cl, \, {1-}$; 
  Nitrate, $(NO3), \, {1-}$;  Carbonate, $(CO3), \, {2-}$; 
  Sulfate, $(SO4), \, {2-}$. 
  } 
 \else{ 
  Vitriol of potash, Sulphat of potash, Potassium sulfate, $K_2SO_4$; 
  or 
  Mephite of barytes, Carbonate of barytes, Barium carbonate, $BaCO_3$;  
  require non-trivial and consecutive diachronic ontological alignments. 
  The ionic valencies (of today's chemistry) involved in the next 
  reactions are as follows;   
  Cations:   
   Hydrogen, $H^{1+}$; Potassium, $K^{1+}$; Ammonium, $(NH_4)^{1+}$; 
   Barium, $Ba^{2+}$; Calcium, $Ca^{2+}$; 
  Anions:  
   Hydroxide, $(OH)^{1-}$;  Chloride, $Cl^{1-}$; Nitrate, $(NO_3)^{1-}$;  
   Carbonate, $(CO_3)^{2-}$; Sulfate, $(SO_4)^{2-}$. 
   }\fi 
  From these ionic valencies it is easy to set up the following 
 stoichiometric balance equations, dry and wet.    

  In contrast to the sophisticated calculations based on the numerical 
 values displayed at Morveau's table and its underlying additive 
 structure, Geoffroy's table, see Figure 9, displays pure and simple 
 ranking orders.     

 \renewcommand{\baselinestretch}{0.82}
 \parskip 0.8mm

  \iflatextortf   
   { \mbox{} \vspace{3mm} 
    \centerline{Affinity table and calculation examples by Morveau.} 
    \centerline{(To be handled ``as'' figures in the editing process)} 
     \vspace{3mm} \mbox{} } 
  \else


 \begin{footnotesize} 

 \begin{table}[h]
 \caption{Guyon de Morveau's Table of Numerical Expression of Affinities} 
 \centering 
 \begin{tabular}{c c c c c c}
 \hline
 Base$\slash$ Acid &
   Vitriolic & Nitric & Muriatic & Acetic & Mephitic \\
 \hline  \hline  
 Barytes  & 65 & 62 & 36 & 28 & 14 \\
 Potash   & 62 & 58 & 32 & 26 &  9 \\
 Soda     & 58 & 50 & 31 & 25 &  8 \\
 Lime     & 54 & 44 & 20 & 19 & 12 \\
 Ammonia  & 46 & 38 & 14 & 20 &  4 \\
 Magnesia & 50 & 40 & 16 & 17 &  6 \\
 Alumina  & 40 & 36 & 10 & 15 &  2 \\
 \hline
 \end{tabular}
 \end{table} 

 \end{footnotesize} 

 \mbox{} \vspace{-3mm}  \mbox{}

 \begin{small}  

\begin{table}[h]  
 \centering 
 \begin{tabular}{ccccccc} 
 & & \multicolumn{3}{c}{$\underbar{
  \mbox{Muriate of Potash,} \ KCl}$} & & 
  \mbox{}\vspace{1mm} \\ 
 Muriate of & \ldelim\{{3}{1mm} & 
  \mbox{Muriat.Ac.} & 32 & \mbox{Potash}  
  & \rdelim\}{3}{1mm} & Mephite \\ 
 Barytes, & & 36 &  +  & 9 (=45) & & of Potash, \\ 
 $BaCl_2$ & & \mbox{Barytes} &  14 (=46) & \mbox{Meph.Ac.} & & 
 $K_2CO_3$ \vspace{-3mm} \\  
  & & \multicolumn{3}{c}{ $\underbrace{\hspace{120pt}}$ } &  \\  
  & & \multicolumn{3}{c}{ \mbox{Mephite of Barytes, \ $BaCO_3$} } &  \\  
 \end{tabular} 
 \end{table}  
 \[ 
    BaCl_2 + K_2CO_3 \rightarrow 2KCl + BaCO_3 \ ,    
 \] 
 \[ 
   [ Ba (OH)_2, 2HCl ] + [ 2KOH, H_2CO_3 ] \rightarrow 
   2[ KOH , HCl ] + [ Ba (OH)_2, H_2CO_3 ] \ ;  
 \] 
 \[ 
    \mbox{Quiescent affinities} = 36 + 9 = 45 < 
    \mbox{Divellent affinities} = 32 + 14 = 46 \ . 
 \]   
 
\end{small}


 \begin{small} 
      
\begin{table}[h]  
 \centering 
 \begin{tabular}{ccccccc} 
 & & \multicolumn{3}{c}{$\underbrace{
  \mbox{Vitriol of Potash,} \ K_2SO_4}$} & & 
  \mbox{}\vspace{1mm} \\ 
 Sulph.of & \ldelim\{{3}{1mm} & 
  \mbox{Vitriol.Ac.} & 62 & \mbox{Potash}  
  & \rdelim\}{3}{1mm} & Mephite \\ 
 Barytes, & & 65 &  +  & 9 (=74) & & of Potash, \\ 
 $BaSO_4$ & & \mbox{Barytes} &  14 (=76) & \mbox{Meph.Ac.} & & 
 $K_2CO_3$ \vspace{-3mm} \\  
  & & \multicolumn{3}{c}{ $\underbrace{\hspace{120pt}}$ } &  \\  
  & & \multicolumn{3}{c}{ \mbox{Mephite of Barytes, \ $BaCO_3$} } &  \\  
 \end{tabular} 
 \end{table}  
 \[ 
    BaSO_4 + K_2CO_3 \rightarrow K_2SO_4 + BaCO_3 \ ,    
 \] 
 \[ 
   [ Ba (OH)_2, H_2SO_4 ] + [ 2KOH, H_2CO_3 ] \rightarrow 
   [ 2KOH, H_2SO_4 ] + [ Ba (OH)_2, H_2CO_3 ] \ ;  
 \] 
 \[ 
    \mbox{Quiescent affinities} = 65 + 9 = 74 < 
    \mbox{Divellent affinities} = 62 + 14 = 76 \ . 
 \] 
 
 \end{small}  


 \begin{small} 
 
\begin{table}[h]  
 \centering 
 \begin{tabular}{ccccccc} 
 & & \multicolumn{3}{c}{$\underbrace{
  \mbox{Vitriol of Lime,} \ CaSO_4}$} & & 
  \mbox{}\vspace{1mm} \\ 
 Sulph.of & \ldelim\{{3}{1mm} & 
  \mbox{Vitriol.Ac.} & 54 & \mbox{Lime}  
  & \rdelim\}{3}{1mm} & Nitrate \\ 
 Ammonia, & & 46 &  +  & 44 (=90) & & of Lime, \\ 
 $(NH_4)_2SO_4$ & & \mbox{Ammonia} &  38 (=92) & \mbox{Nitr.Ac.} & & 
 $Ca(NO_3)_2$ \vspace{-3mm} \\  
  & & \multicolumn{3}{c}{ $\underbrace{\hspace{120pt}}$ } &  \\  
  & & \multicolumn{3}{c}{ \mbox{Nitrate of Ammonia, \ $NH_4NO_3$} } &  \\  
 \end{tabular} 
 \end{table}  
 \[ 
    (NH_4)_2SO_4 + Ca(NO_3)_2 \rightarrow CaSO_4 + 2NH_4NO_3 \ ,    
 \] 
 \[ 
   [ 2NH_4OH, H_2SO_4 ] + [ Ca(OH)_2, 2HNO_3  ] \rightarrow 
   [ Ca(OH)_2, H_2SO_4 ] + [ 2NH_4OH, 2HNO_3 ] \ ;  
 \] 
 \[ 
    \mbox{Quiescent affinities} = 46 + 44 = 90 < 
    \mbox{Divellent affinities} = 54 + 38 = 92 \ . 
 \]

 \end{small}  

 \mbox{}

 \fi 

  \begin{small} 

  \begin{figure}[bh] 
  \iflatextortf
   {} 
  \else   
   \centerline{  
     \includegraphics[  width=5.5in,  
    trim=0.0in 0.0in 0.0in 0.0in, clip]{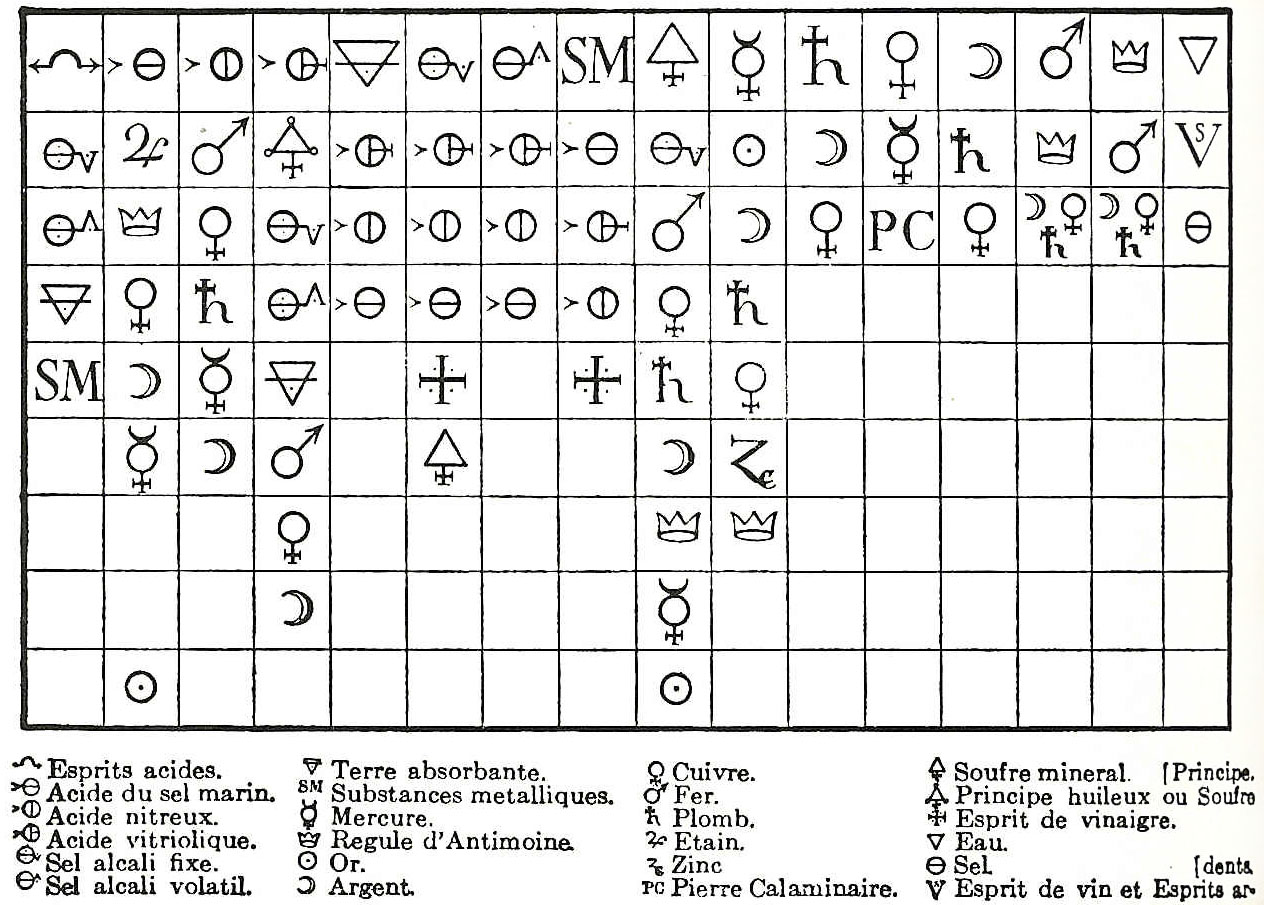} 
     \mbox{} \ \mbox{} 
    } 
  \fi  
  \vspace{1mm} 
  \centerline{Figure 9: Affinity table by Geoffroy (1718).} 
  \end{figure} 

 \end{small} 
  



 \end{document}